\documentclass{elsart}

\usepackage{epsfig}
\usepackage{amssymb,amsmath}

\newcommand{\MAC}[0]{{\sc mac}}

\begin{document}

\begin{frontmatter}

\title{Distributive routing \& congestion control \\
       in wireless multihop ad hoc \\
       communication networks}

\author[label1,label2,label3]{Ingmar Glauche}
  \ead{ingmar.glauche@imise.uni-leipzig.de}
\author[label1,label4]{Wolfram Krause}
  \ead{krause@th.physik.uni-frankfurt.de}
\author[label1]{Rudolf Sollacher}
  \ead{rudolf.sollacher@siemens.com}
\author[label1]{Martin Greiner}
  \ead{martin.greiner@siemens.com}
\address[label1]{Corporate Technology, Information \& Communications, 
                 Siemens AG, 
                 D-81730 M\"unchen, Germany}
\address[label2]{Institut f\"ur Theoretische Physik,
                 Technische Universit\"at Dresden,
                 D-01062 Dresden, Germany}
\address[label3]{Institut f\"ur Medizinische Informatik, Statistik
                 und Epidemiologie,
                 Universit\"at Leipzig,
                 Liebigstr.\ 27,
                 D-04103 Leipzig, Germany}
\address[label4]{Institut f\"ur Theoretische Physik,
                 Johann Wolfgang Goethe-Universit\"at Frankfurt,
                 Postfach 11 19 32,
                 D-60054 Frankfurt am Main, Germany}

\begin{abstract}
Due to their inherent complexity, engineered wireless multihop ad hoc
communication networks represent a technological challenge. Having no
mastering infrastructure the nodes have to selforganize themselves in
such a way that for example network connectivity, good data traffic
performance and robustness are guaranteed. In this contribution the
focus is on routing \& congestion control. First, random data traffic
along shortest path routes is studied by simulations as well as
theoretical modeling. Measures of congestion like end-to-end time
delay and relaxation times are given. A scaling law of the average
time delay with respect to network size is revealed and found to
depend on the underlying network topology. In the second step, a
distributive routing \& congestion control is proposed. Each node
locally propagates its routing cost estimates and information about
its congestion state to its neighbors, which then update their
respective cost estimates. This allows for a flexible adaptation of
end-to-end routes to the overall congestion state of the network.
Compared to shortest-path routing, the critical network load is
significantly increased.
\end{abstract}


\end{frontmatter}

\newpage
\section{Introduction}
\label{sec:intro}

In two previous Papers \cite{gla03a,kra04} we have already discussed
so-called wireless multihop ad hoc networks. They represent an engineered 
communication network, which reveals many facets of very intriguing and 
complex behavior. In this respect they fit nicely into the 
cross-disciplinary realm of the Statistical Physics of complex networks 
\cite{alb02,dor03,new03}, which has already opened its doors for other 
communication networks like the Internet, but also for biological and 
social networks.

Wireless multihop ad hoc networks represent an infrastructureless 
peer-to-peer generalization of todays wireless cellular phone networks. 
Instead of being slaved to a central control authority, each node not 
only sends or receives packets, but also forwards them for others.
Consequently, communication packets hop via inbetween ad hoc nodes to 
connect the initial sender to the final recipient. A lot of coordination 
amongst the nodes is needed for the overall network to perform well. They 
have to ensure network connectivity, good data-traffic performance and 
robustness against various forms of perturbations, just to name but the 
most important issues. Because of this intrinsic coordination, wireless
multihop ad hoc networks represent an excellent example of what is 
called a selforganizing network. However, their biggest challenge is 
yet to come, how to get selforganization to work.

The connectivity issue has already been discussed quite extensively
\cite{gla03a,gup98,bet02,dou02,xue04}, also addressing interference 
effects \cite{dou03,sol04}. In one form or the other all these efforts
relate to continuum percolation \cite{mee96,sto95,dal02}. An interesting
distributive scheme has been put forward in \cite{gla03a}, which turned
out to be amazingly robust, guaranteeing strong network connectivity
almost surely; we will briefly touch upon this scheme again in Sect.\
\ref{subsec:twoA}. 
--
The robustness issue with respect to selfish users has received
inspirations from the biological immune system and distributive
algorithmic suggestions have been put forward \cite{bou04}.

As to data-traffic performance, estimates on the throughput, i.e.\ the 
capacity of how much end-to-end traffic the network is able to handle
without overloading, have been given. In \cite{gup00} a rigorous upper 
bound has been derived to scale with the square root of the network 
size. Refined estimates have been given in \cite{kra04}, revealing that 
the scalability of the throughput depends on the underlying network 
structure. Besides several other idealistic assumptions, these 
estimates have employed shortest-path routing. Although several 
proactive and reactive routing schemes have already been discussed 
\cite{manet,nistb,mob02,mob03}, we are not aware of any selfoptimizing 
scheme, which also accounts for congestion avoidance. 

In this Paper we will propose a prototype for such a highly wanted 
distributive and adaptive routing \& congestion control. The idea is
that every node keeps an estimate of how much it costs to send packets
to final destinations and to update these estimates in a distributive
manner. The latter can be achieved in a very elegant way without any
additional exchange of control information. Whenever a node is 
actively involved in a forwarding one-hop transmission, it silences 
its neighbors anyhow, so that those do not interfere on the shared 
wireless propagation medium. This blocking is called medium access 
control. Upon blocking its neighbors, the node is able to distribute 
its routing cost estimates and its congestion state to them, which 
those then use to update their cost estimates. In the technical jargon 
of engineers, this distributive scheme corresponds to a coupling of 
the medium access control layer with the routing layer.

The structure of this Paper is as follows. Sect.\ \ref{sec:two} 
summarizes the key operational features of wireless multihop ad hoc
networks, introduces plausible simplifications and describes the setup
of generic simulations with random data traffic. Shortest-path routing
is used in Sect.\ \ref{sec:three} to investigate certain fingerprints
of congestion like end-to-end time delay and single-node relaxation
times. Whenever possible, the numerical simulations are accompanied 
with analytic modeling. Sect.\ \ref{sec:four} presents the details of 
the proposed distributive routing \& congestion control and compares its 
results to those obtained with the shortest-path routing. The conclusion 
and a short outlook are given in Sect.\ \ref{sec:conclusion}.

\section{Some basics on wireless multihop ad hoc networks}
\label{sec:two}

We explain the key features of wireless multihop ad hoc networks
hand in hand with some simplifications.

\subsection{Geometric ad hoc graphs}
\label{subsec:twoA}

The first simplification is to neglect mobility and to distribute $N$
nodes onto the unit square in a random and homogeneous way. Then,
according to a simple isotropic propagation-receiver model, a
unidirectional link from node $i$ to node $j$ exists, if 
\begin{equation}
\label{eq:propagation}
  \frac{P_{i}/R_{ij}^\alpha}
       {\mbox{\sc noise} + 
        \sum_{\mathrm{active}k} P_{k}/R_{kj}^\alpha} 
    \geq  {\mbox{\sc snr}}
          \; .
\end{equation}
$P_{i}$ denotes the transmission power of node $i$ and $R_{ij}$
represents the Euclidean distance between $i$ and $j$. The path-loss
exponent $\alpha$ is assumed to be constant.  Without any loss of
generality the variables $\mbox{\sc noise}$ and $\mbox{\sc snr}$ are set 
equal to one. Condition \eqref{eq:propagation} guarantees that $j$ is 
able to listen to $i$, i.e.\ $i{\rightarrow}j$. Throughout this Paper we
will neglect the interference sum over other active nodes $k$ in the
denominator of \eqref{eq:propagation}; this is justified once $\alpha$
is not too close to $2$ \cite{sol04}.

With these simplifications, wireless communication networks can be
modeled as graphs $\mathcal{G} = (\mathcal{N},\mathcal{L})$, where
$\mathcal{N}$ refers to the set of nodes and $\mathcal{L}$ to the set
of links. In general these links in $\mathcal{L}$ are directional
links $i\rightarrow j$. The subset $\mathcal{L}^{\rm
bidir}\subset\mathcal{L}$ represents the complete set of all
bidirectional links $i\leftrightarrow j$. Although not strictly
required, bidirectional links are preferred for the operation of
wireless ad hoc networks because many communication protocols require
instant feedback. The subset $\mathcal{N}_{i}\subset\mathcal{N}$ is
called the communication neighbourhood of node $i$ and represents the
complete set of nodes $j\in\mathcal{N}_{i}$ that all have
bidirectional links $j\leftrightarrow i$ in $\mathcal{L}^{\rm bidir}$
with node $i$. The node degree $k_{i}$ of node $i$ is the number of
nodes contained in $\mathcal{N}_{i}$. In a similar fashion
$\mathcal{N}^{\rm out}_{i}$ defines the set of nodes that have at
least an unidirectional link $i\rightarrow j$ from $i$. A
communication route or path is a sequence of nodes such that there are
bidirectional links in $\mathcal{L}^{\rm bidir}$ between all
consecutive pairs of nodes. A shortest path between two nodes
$i,f\in\mathcal{N}$ is a route containing fewest possible number of
nodes. The average length of all shortest paths over all node pairs
$i$, $f$ is refered to as the diameter $D$ of the network.

One further step is needed in order to fully specify wireless multihop
ad hoc network graphs: assignment of the transmission power $P_i$ for
all nodes. Most widley used is the constant transmission power rule, 
where the same transmission power $P_{i}=P$ is assigned to all nodes 
$i \in \mathcal{N}$ \cite{gup98,dou02,gup00}. All existing links are then 
bidirectional. Once the transmission power is chosen such that
$P=(k_\mathrm{target}/\pi{N})^{\alpha/2}$, an average node will have 
$k_\mathrm{target}$ bidirectional neighbors. This target degree has to be 
$k_\mathrm{target}\geq 4.52$ for a connected giant component to exist
independent of the network size \cite{dal02}, but for the entire 
network to be strongly connected it needs to be larger 
\cite{gla03a,bet02,xue04}. We adopt the value $k_\mathrm{target}=24$, 
which guarantees strong network connectivity almost surely for 
network sizes up to several thousands \cite{gla03a}. We will call 
wireless multihop ad hoc network graphs generated with this power
assignment as const-$P$ networks.

A different power assignment has been presented in Ref.\ \cite{gla03a},
which is more energy efficient. It is based on a distributive 
assignment. In a nutshell, each node $i$ forces the $k_\mathrm{min}$ 
closest nodes $j$ to adjust their transmission powers to 
$P_j=R_{ij}^\alpha$, while adopting the value $P_i=\sup_j P_j$ for 
itself. Its own value can be increased further whenever another 
close-by node forces $i$ in return to have an even larger transmission 
power. In this respect each node has at least $k_\mathrm{min}$ 
bidirectional neighbors. We adopt the value $k_\mathrm{min}=8$, which 
guarantees strong network connectivity almost surely for network sizes 
up to several thousand nodes \cite{gla03a}. We will call wireless multihop 
ad hoc network graphs generated with this heterogeneous power 
assignment as minimum-node-degree networks. 

\begin{figure}
\begin{center}
\epsfig{file=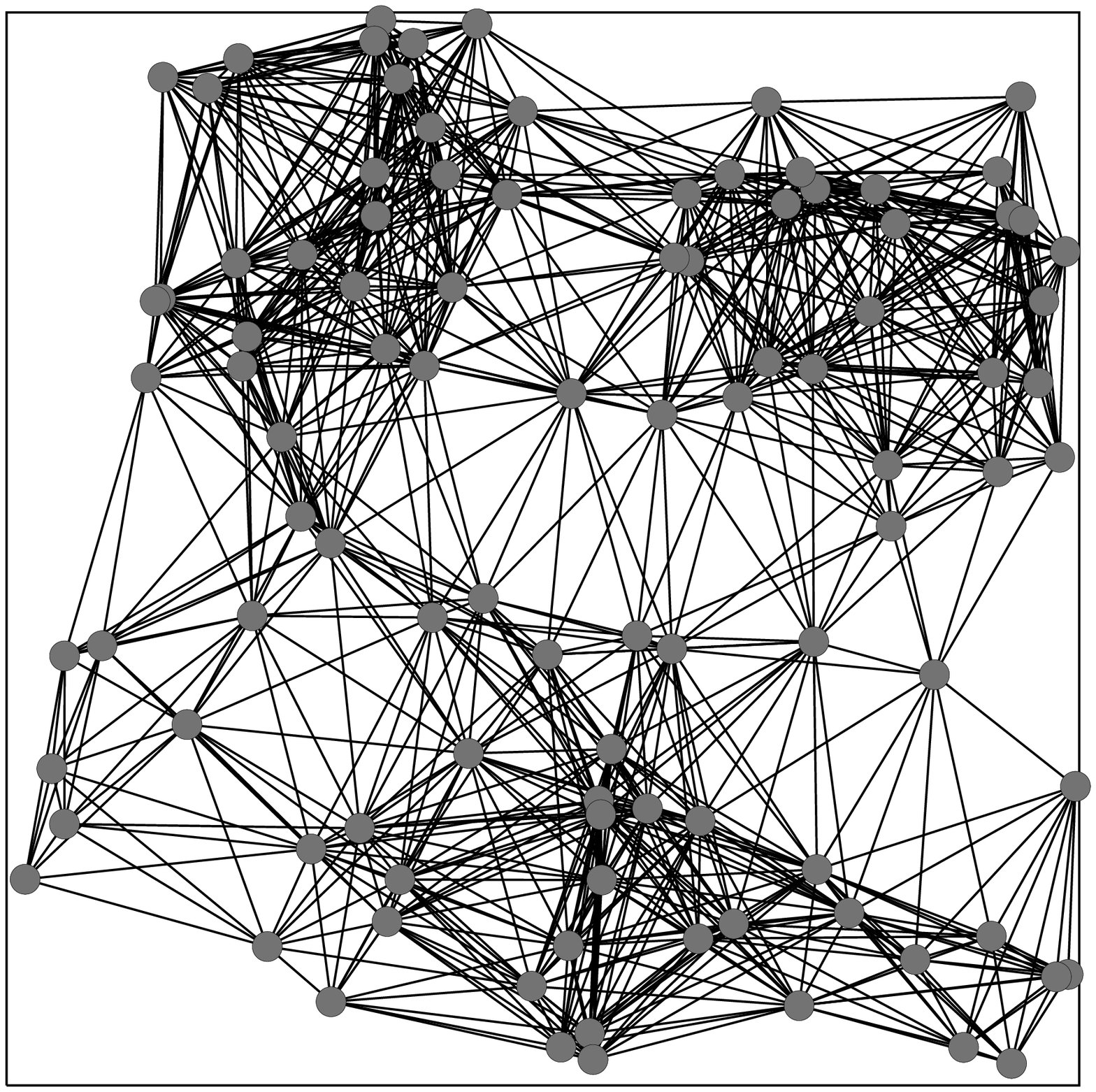,width=0.49\textwidth}
\epsfig{file=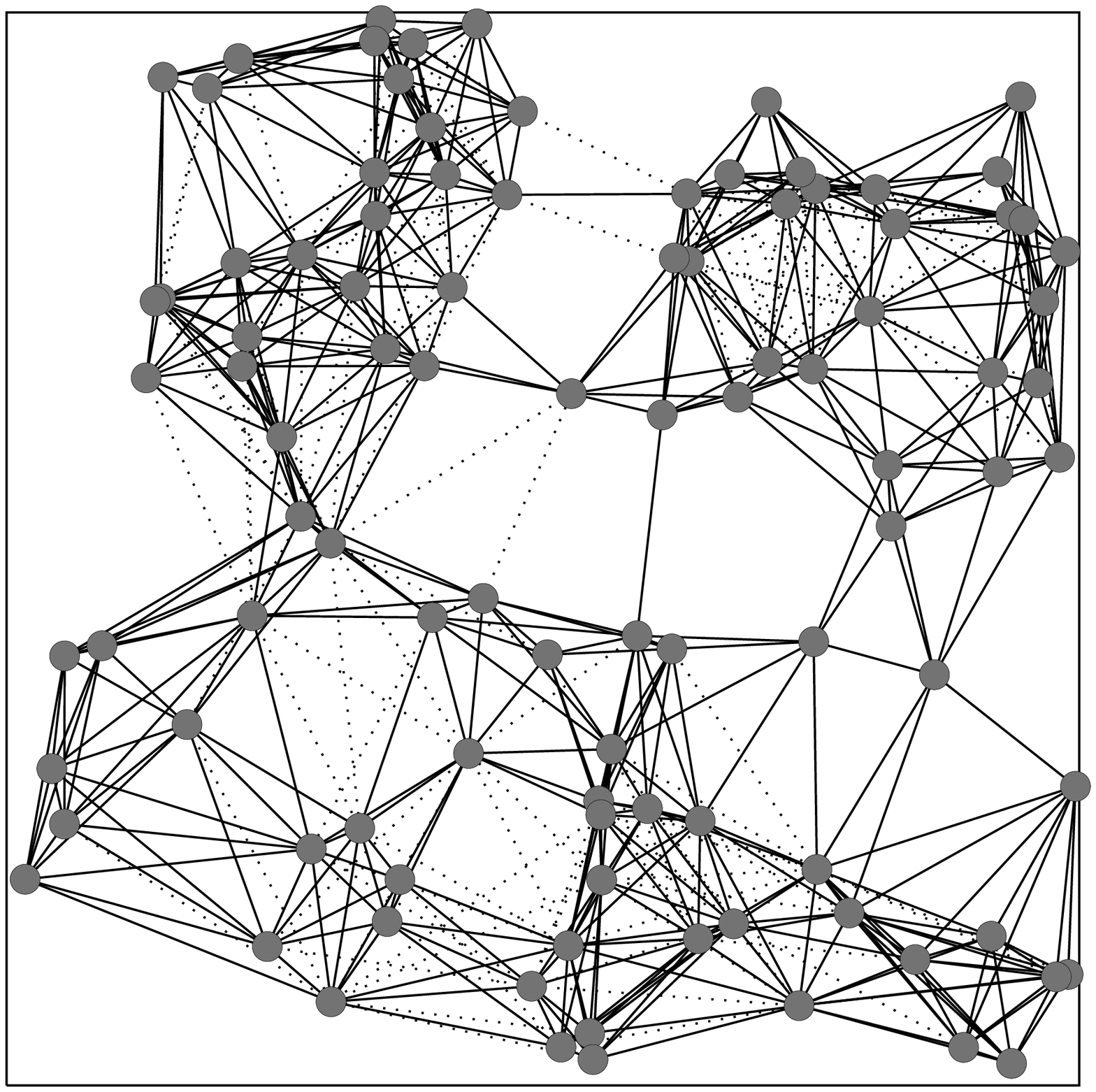,width=0.49\textwidth}
\caption{
Random geometric wireless multihop ad hoc graphs obtained with (left) 
const-$P$ and (right) minimum-node-degree transmission power assignment.
$N=100$ nodes have been randomly and homogeneously distributed onto a
unit square. Solid/dotted links are bi-/unidirectional.
}
\label{fig:pattern}
\end{center}
\end{figure}

Fig.\ \ref{fig:pattern} illustrates two random geometric graphs, one 
obtained with the const-$P$ assignment and the other with the 
minimum-node-degree assignment. We explicitely point out that the 
existence of links in $\mathcal{L}$ is a direct result of the spatial 
positions and the transmission powers of the nodes. In that sense not 
every possible set of links $\mathcal{L}$ can be realized by an 
appropriate power assignment. This is in clear contrast to any wired 
network where such restrictions do not influence the existence of 
connections between different nodes.

\subsection{Generic data traffic}
\label{subsec:twoB}

In order to study the statistical properties of data traffic on
wireless multihop ad hoc networks, the generic simulation model as
already presented in Ref.\ \cite{kra04} has been applied. For the sake
of illustration of the key mechanisms, we now give again a short
outline.  The simulation is based on discrete time steps. At the very
beginning of a time step a new data packet of fixed size can be
generated at each node $i \in \mathcal{N}$ with a probability $\mu_{i}
= \mu < 1$, which is also referred to as the packet creation rate. In
case of creation at a certain node $i$, a destination $f$ is randomly
chosen among $\{\mathcal{N} \setminus i\}$ and the packet is put at
the end of $i$'s buffer queue, assumed to have infinite
capacity. Nodes, for which a new packet has been created, are blocked
and are not involved in any further communication action for the
remainder of this time step.  During a short contention phase
following the packet creation phase the non-blocked nodes with a
non-zero queue compete for gaining sender status. A competing node $i$
is randomly picked first and obtains permission to transmit its
first-in-line packet. It then makes a decision, to which neighbor
$j\in\mathcal{N}_i$ the packet with final destination $f$ is
forwarded. In its simplest form, this could be shortest-path routing
or, in a more sophisticated form, routing depending on the congestion
state. In order to reduce mutual interference within the shared
communication medium, the sending as well as receiving node block
their respective outgoing neighbors $\{ (\mathcal{N}^{\rm out}_{i}
\setminus j ) \cup (\mathcal{N}^{\rm out}_{j} \setminus i) \}$ for the
remainder of this time step; this blocking is called medium access
control (\MAC). Only then another node with non-zero queue that has
not been blocked so far is chosen at random to attempt the
transmission of its first-in-line packet. If the intended receiver has
already been blocked before, the node tests its second-in-line packet
and so on, until either the first idle recipient is found or the end
of its queue is reached. This service discipline is denoted as
first-in-first-possible-out. If this node succeeds to gain sender
status, it then {\MAC-}blocks again its remaining outgoing neighbors
as well as those of the receiving node. This iteration is repeated
until no free one-hop transmissions are left. Finally, all nodes with
sending permission then submit their selected packet and remove it
from their queue. The receiving nodes either add the incoming packet
to the end of their queue or, if they are the final recipient, destroy
the packet.

\section{Fixed shortest-path routing: properties of data traffic}
\label{sec:three}

A particular simple form of routing uses shortest paths with respect to
the hop-metric. For the network to learn about the shortest routes all
by itself, each node is required to flood discovery information into the
whole network, collect the feedback and store those routes with the
shortest hop distance. In principle degeneracy might occur, which would 
allow to pick the least-congested shortest path between initial sender 
$i$ and final recipient $f$ as a modest form of congestion control; we 
will come back to this in Sect.\ \ref{subsec:fourA}. For all of this 
Section we prefer to discard degeneracy, pick one of the shortest 
degenerate paths at random and forward all packets originating in $i$ 
and destined for $f$ along it. This restriction allows for several 
analytical insights and points to the specific needs for improvement 
which a more sophisticated routing \& congestion control has to take
care of.

\subsection{End-to-end time delay I: simulation results}
\label{subsec:threeA}

We define the end-to-end (e2e) time delay of a packet to be the number 
of time steps between the generation at the originating node and the 
destruction at the final receiver. Its temporal and network-ensemble 
average provides a measure of the network performance. The direct 
sampling of end-to-end times within the generic data traffic simulations 
implies a tagging of each packet with its creation time and is needed to 
extract respective distributions. Although average e2e-time delays can 
also be determined via this route, an indirect sampling procedure is 
more efficient in that case. In the subcritical regime 
$\mu < \mu_{crit}$ the average number $\mu N$ of packets created within 
the overall network per time step must be equal to the number of packets 
delivered per unit time. Since the average time a packet spends in the 
network is $\langle t_{e2e} \rangle$, we can assume that 
$\langle N_{active} \rangle / \langle t_{e2e} \rangle$ packets are 
delivered to their final destination per unit time, where 
$\langle N_{active} \rangle = \langle \sum_i n_i \rangle$ represents the 
average total number of active packets after a stationary network state
has been reached. This leads to Little's Law,
\begin{equation}
\label{eq:littlelaw}
  \frac{\langle N_{active} \rangle}{\langle t_{e2e} \rangle} 
    =  \mu N 
       \; ,
\end{equation}
well known in queuing theory \cite{gel98}. Since the network size $N$
and the packet creation rate $\mu$ are known to us and 
$\langle N_{active} \rangle$ is easily sampled, Little's Law allows 
for the indirect determination of the average end-to-end time delay.

\begin{figure}
\begin{centering}
\epsfig{file=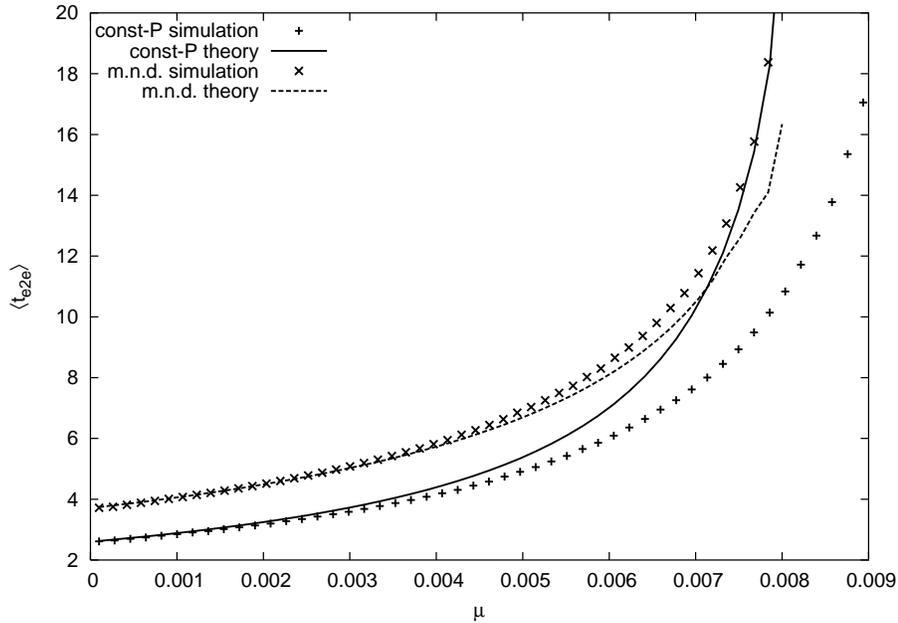,width=12cm}
\caption{
Sample- and node-averaged end-to-end time delay $t_{e2e}(\mu)$ 
determined from generic data traffic simulations (symboled curves) and 
the analytic estimate using Eqs.\ (\ref{eq:littlelaw}), 
(\ref{eq:avgni}), (\ref{eq:muin}) and (\ref{eq:muout}) (curves 
without symbols). The network size has been fixed to $N=100$. The two
transmission power assignments are const-$P$ (vertical crosses) with 
$k_\mathrm{target} = 24$ and minimum-node-degree (rotated crosses) with 
$k_{min}=8$.
}
\label{fig:te2e}
\end{centering}
\end{figure}

Fig.\ \ref{fig:te2e} illustrates the average e2e-time delay as a 
function of the packet creation rate $\mu$. As expected, it increases 
with the network load $\mu$ and diverges at the critical packet 
creation rate $\mu_\mathrm{crit}$, which is a clear 
phase-transition-like sign that the system is about to leave its 
uncongested subcritical state and to enter its congested supercritical
state for $\mu>\mu_\mathrm{crit}$. We also observe that 
$\mu_\mathrm{crit}$ is different for the const-$P$ and the 
minimum-node-degree networks; see Ref.\ \cite{kra04} for more details 
on the dependence of the end-to-end throughput $\mu_\mathrm{crit}N$ on 
the underlying network structure. In the limit $\mu\to 0$ the average 
e2e-time delay converges to the network diameter 
$\langle t_{e2e} \rangle \to D$. Although this observation is 
intuitively clear, we will give an analytic support in the next 
Subsection. Due to the differing network diameter, it appears that with 
respect to e2e-time delay the const-$P$ networks perform better than 
the minimum-node-degree networks. Note however, that their respective 
parameters $k_\mathrm{target}$ and $k_\mathrm{min}$ have been chosen 
from the connectivity perspective only. A larger $k_\mathrm{min}$
would lower the network diameter for the minimum-node-degree networks.
In Sect.\ \ref{subsec:threeC} we will present another form of 
performance comparison between the two network models.

\begin{figure}
\begin{center}
\epsfig{file=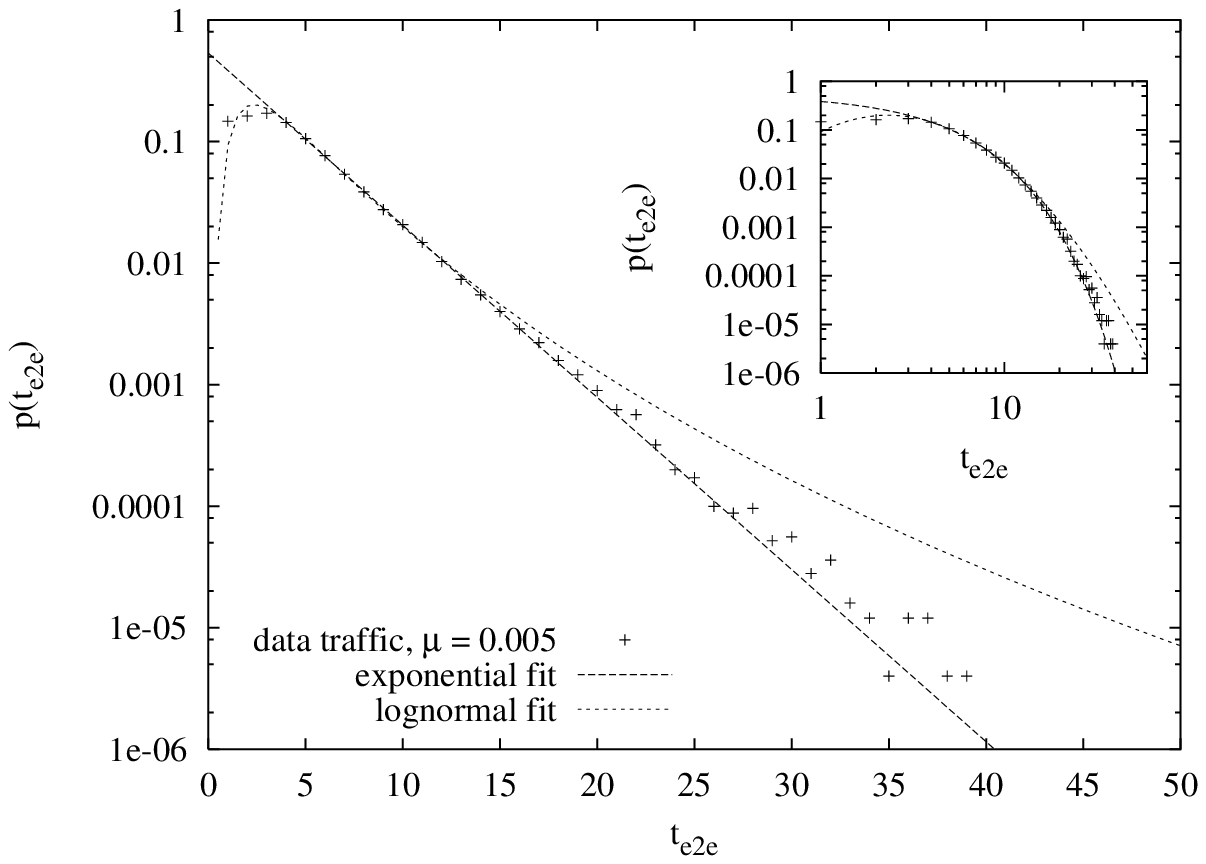,width=0.8\textwidth}
\epsfig{file=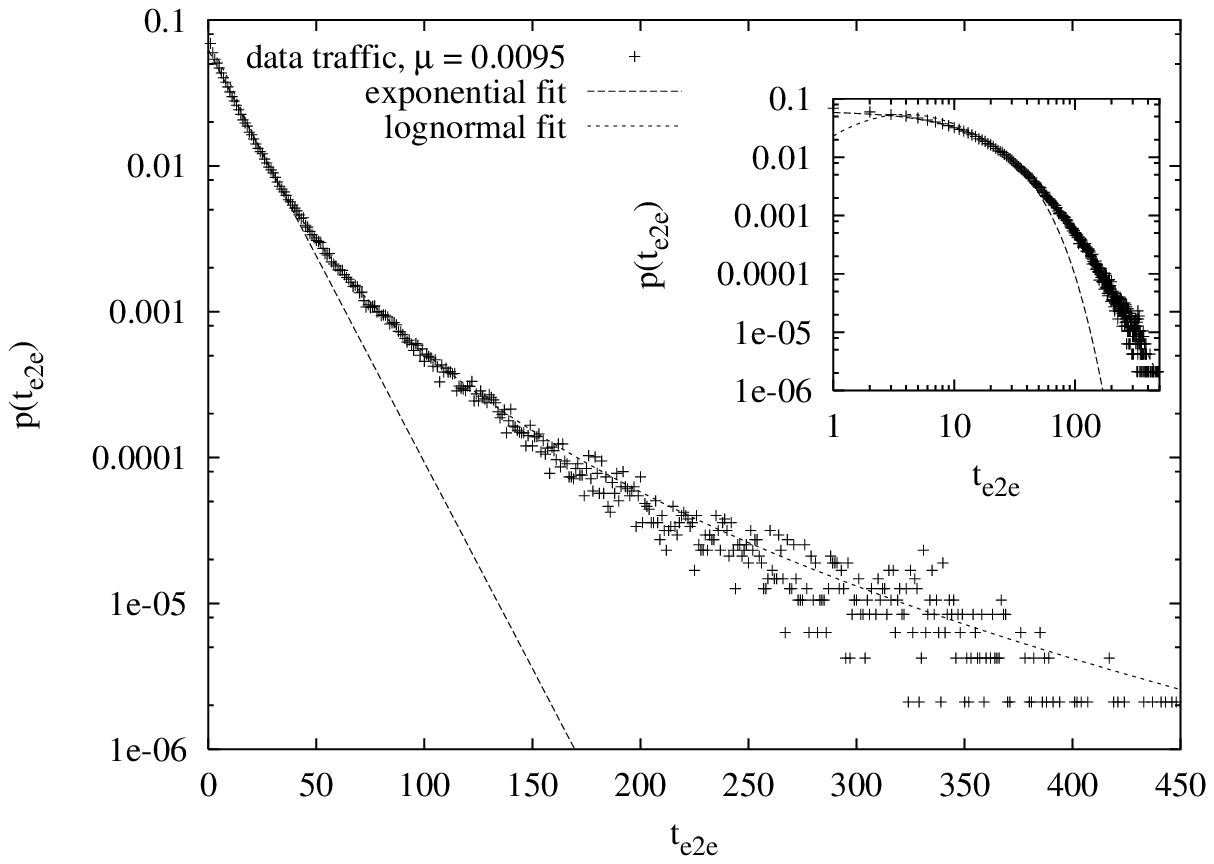,width=0.8\textwidth}
\caption{
Distribution of the end-to-end time delay obtained from a generic data
traffic simulation with fixed shortest-path routing on a typical 
const-$P$ network with $N=100$ nodes. (Top) $\mu = 0.005$ well below
and (bottom) $\mu = 0.0095$ close to the critical packet creation rate
$\mu_\mathrm{crit}$. Best fits with an exponential and a log-normal
distribution are also shown. The insets represent log-log plots.
}
\label{fig:constP24nodes100macB.delaytimes_fit}
\end{center}
\end{figure}

The distribution $p(t_{\rm e2e})$ of the e2e-time delay obtained from 
the generic data-traffic simulation with fixed shortest-path routing 
is shown in Fig.\ \ref{fig:constP24nodes100macB.delaytimes_fit}. It is 
a network-wide distribution, which has been sampled over all generated 
packets. The employed const-$P$ network realization consists of 
$N=100$ nodes. Safely inside the subcritical phase ($\mu = 0.005$) the 
sampled distribution can be fitted well with a generalized exponential 
\begin{equation}
\label{eq:expDistr}
  p_\mathrm{exponential}(t_\mathrm{e2e}) 
    =  \frac{1}{b}e^{-(t_\mathrm{e2e}-a)/b} 
       \; .
\end{equation}
However, for packet creation rates ($\mu = 0.0095$) close to
$\mu^{\rm crit} = 0.0101$ of the particular used network realization a 
log-normal distribution
\begin{equation}
\label{eq:lognormDistr}
  p_\mathrm{lognormal}(t_\mathrm{e2e}) 
    =  \frac{1}{\sqrt{2\pi}\sigma t_\mathrm{e2e}} \, 
       e^{-(\ln t_\mathrm{e2e} - \tau)^2 / (2\sigma^2)}
\end{equation}
provides a better fit. The emergence of a log-normal distribution close
to the critical packet creation rate appears to be an inherent feature 
of communication networks \cite{hub97,sol01}, when nodes are
strongly and collectively coupled via heavy data traffic. 

The tendency to have a small but non-negligible number of packets with
a rather high end-to-end time delay is an undesirable feature for
communication networks. A suppression of this tail in the distribution 
of the end-to-end time delay is regarded as one goal of any advanced 
routing \& congestion control. Of course, another goal is to 
increase the critical packet creation rate.

\subsection{End-to-end time delay II: analytic estimate}
\label{subsec:threeB}

In this Subsection we give an analytic estimate of the mean end-to-end
time delay. The starting point is Little's Law (\ref{eq:littlelaw}), 
which demands to model the average number $N_\mathrm{active}$ of active 
data packets traveling on the network. Upon assuming inter-node 
correlations to be small, we adopt the single-node picture and reduce 
$N_\mathrm{active}=\sum_i n_i$ to the modeling of a single-node queue 
length $n_i$. The latter itself depends on the in- and out-flux rates 
$\mu_i^\mathrm{in}$ and $\mu_i^\mathrm{out}$ of data packets to node 
$i$. If this turns out to be the only dependence, then the probability 
for node $i$ to have $n_i$ packets in its queue can be described by the 
rate equation
\begin{equation}
\label{eq:bufferDistr1}
  p(n_i,t+1) 
    =  \mu^\mathrm{out}_i p(n_i+1,t) 
       + (1-\mu^\mathrm{in}_i-\mu^\mathrm{out}_i) p(n_i,t)
       + \mu^\mathrm{in}_i p(n_i-1,t)
       \; .
\end{equation}
Reducing to the stationary limit $p(n_i,t+1) = p(n_i,t)$ and taking the 
boundary condition $p(n_i<0)=0$ into account, its normalized solution is
\begin{equation}
\label{eq:bufferDistrFinal}
  p(n_i) 
    =  \biggl( 
       \frac{\mu^\mathrm{in}_i}{\mu^\mathrm{out}_i} 
       \biggr)^{n_{i}} 
       \biggl(
       1 - \frac{\mu^\mathrm{in}_i}{\mu^\mathrm{out}_i}
       \biggr)
       \; .
\end{equation}
As a direct consequence, the relations
\begin{equation}
\label{eq:avgni}
  \langle n_i \rangle
    =  \frac{\mu^\mathrm{in}_i/\mu^\mathrm{out}_i}
            {1-\mu^\mathrm{in}_i/\mu^\mathrm{out}_i}
\end{equation}
and 
\begin{equation}
\label{eq:nigeone}
  p(n_i\geq 1) 
    =  \mu^\mathrm{in}_i/\mu^\mathrm{out}_i   
\end{equation}
are derived, which will be of later use.

\begin{figure}
\begin{centering}
\epsfig{file=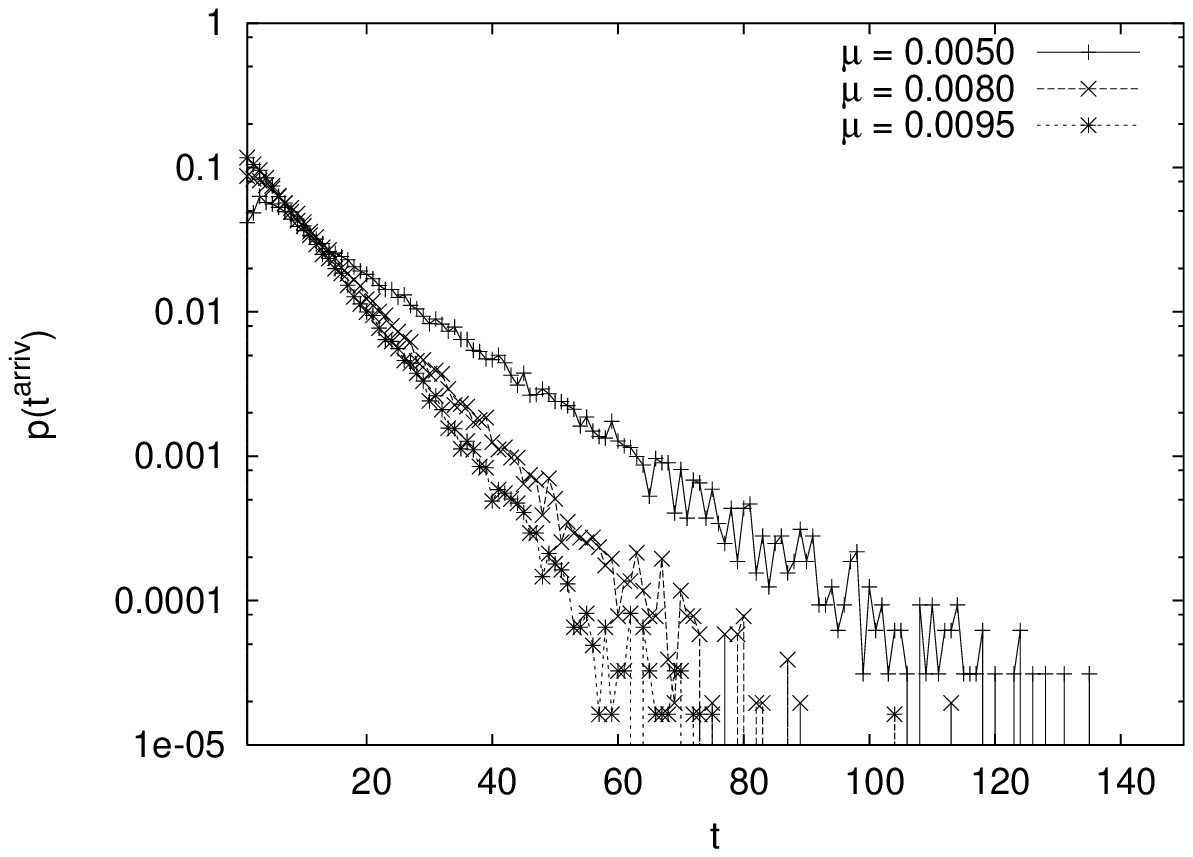,width=9cm}
\epsfig{file=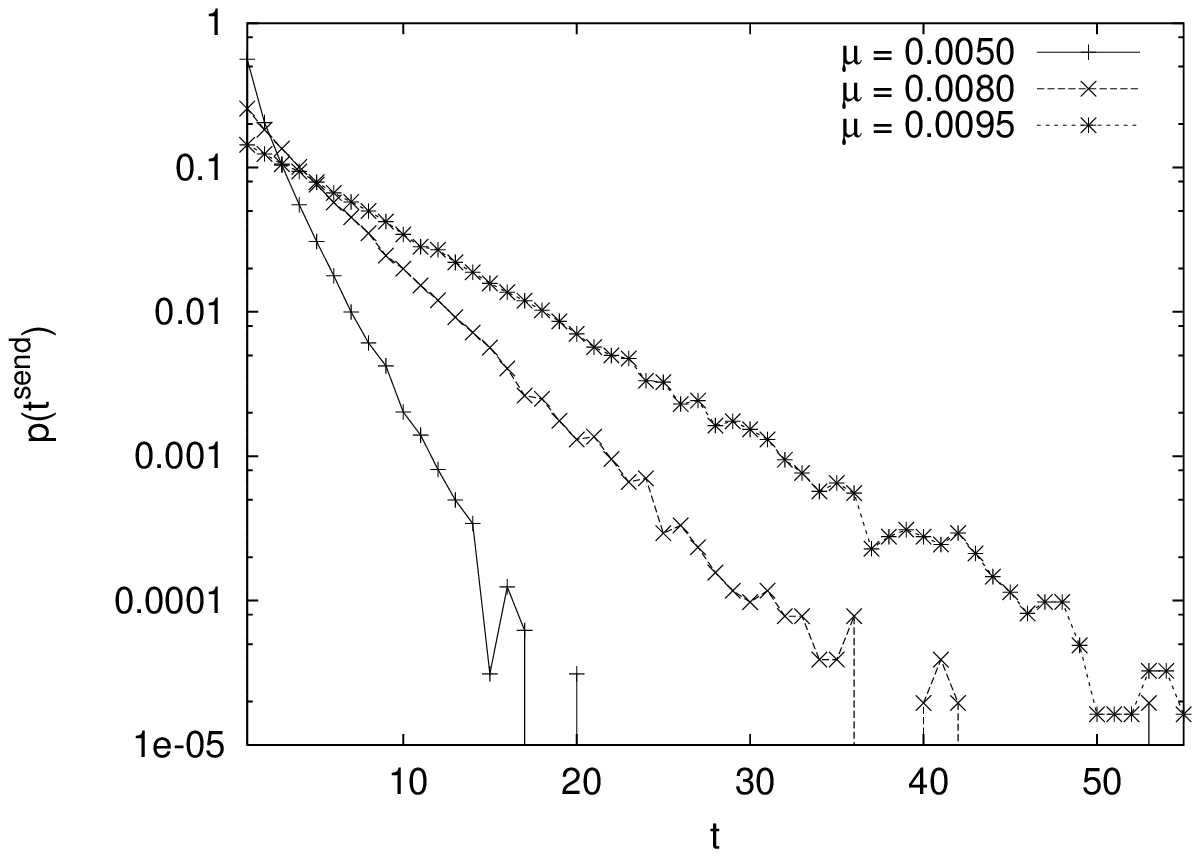,width=9cm}
\epsfig{file=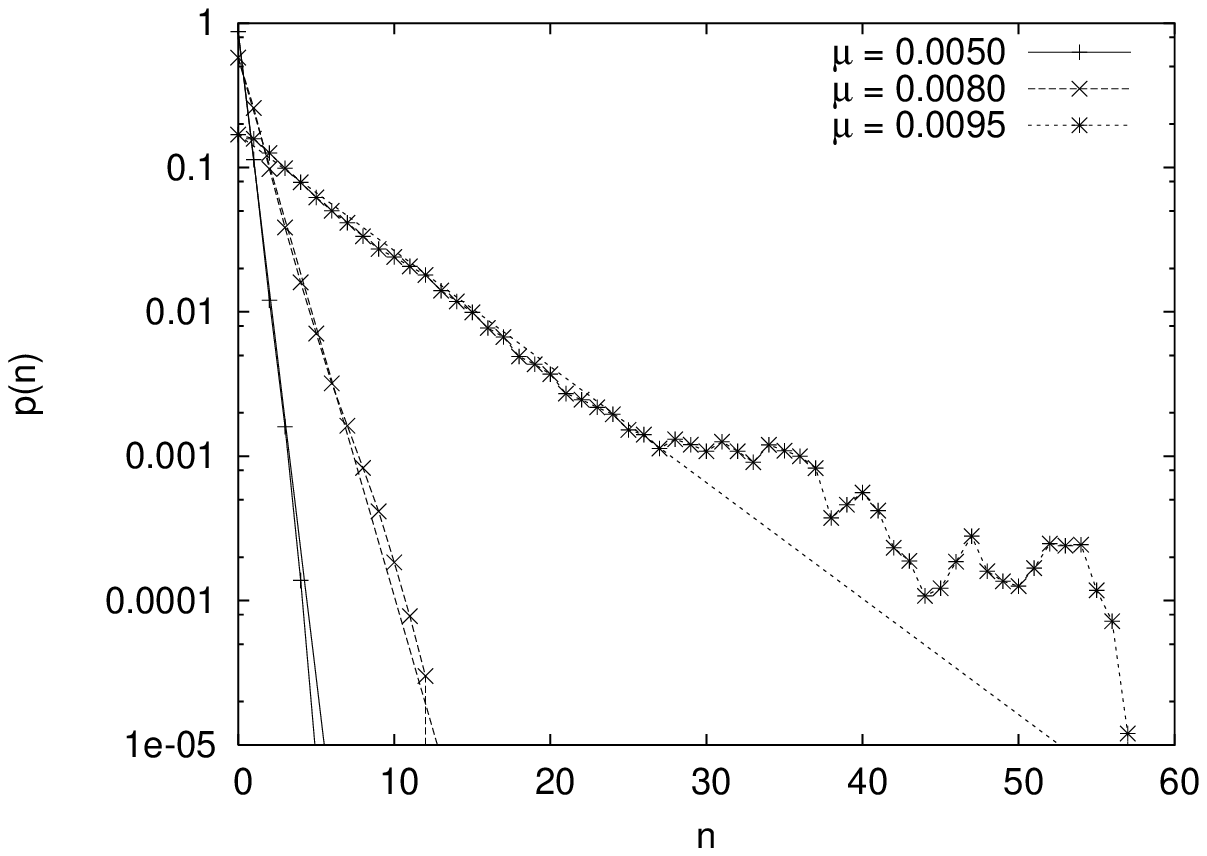,width=9cm}
\caption{
Distributions for (top) interarrival times, (middle) sending times and 
(bottom) queue lengths as observed at the most-critical node of a 
const-$P$ network realization with $N=100$. All shown packet creation 
rates belong to the subcritical phase. The thin lines in (bottom) 
represent the expression (\ref{eq:bufferDistrFinal}), where according to 
(\ref{eq:interarrivalDistr}) the in- and out-flux rate have been taken 
from (top) and (middle) as the reciprocal of the mean interarrival and 
sending times.
}
\label{fig:histo}
\end{centering}
\end{figure}

We have carefully checked the assumption going into 
(\ref{eq:bufferDistr1}); see also \cite{gla03b}. If the queue length
distribution only depends on the in- and out-flux rates, which is filed
as case $M/M/1$ in queuing theory \cite{gel98}, then the interarrival
and sending times statistics should both obey the geometric distribution
\begin{equation}
\label{eq:interarrivalDistr}
  p\left(t^\mathrm{arrive/send}_i=t\right) 
    =  \left(1-\mu^\mathrm{in/out}_i\right)^{t-1} \mu^\mathrm{in/out}_i
       \; ,
\end{equation}
which comes with mean 
$\langle t^\mathrm{arrive/send}_i \rangle = 1/\mu^\mathrm{in/out}_i$ and
reflects the independence of subsequent packet arrival and departure
events. The interarrival time $t^\mathrm{arrive}_i$ is defined as the 
time between two successive arrivals of data packets that are put at the
end of node $i$'s queue. The sending time $t^\mathrm{send}_i$ is defined 
as the time between two successive sending events from node $i$ to any 
of its neighboring nodes $j\in\mathcal{N}_i$ given that the queue is 
non-empty. Fig.\ \ref{fig:histo} illustrates results obtained from the 
generic data traffic simulation, which show that, when focusing on a 
single node of an arbitrary network realization, the distributions of 
interarrival and sending time nicely follow the parameterization 
(\ref{eq:interarrivalDistr}) for various packet creation rates and that 
the extracted in- and out-flux rates lead to a good agreement between 
the queue length distribution (\ref{eq:bufferDistrFinal}) and its 
simulated counterpart.

The modeling is now open for the in- and out-flux rates. We briefly 
outline the results, which have already been derived in a previous Paper 
\cite{kra04}. The in-flux rate
\begin{equation}
\label{eq:muin}
  \mu^\mathrm{in}_i
    =  \frac{\mu B_i}{N-1}
\end{equation}
is proportional to the rate $\mu N$ of newly created packets, of which 
the fraction $B_i/N(N-1)$ will be routed via node $i$ during later time 
steps. The node inbetweeness $B_i$ and, equivalently, the link 
inbetweeness $B_{i{\leftrightarrow}j}$ count the number of shortest 
paths, which go over node $i$ and link $i{\leftrightarrow}j$,
respectively. The modeling of the out-flux rate 
$\mu^\mathrm{out}_i=1/\tau_i$ is equivalent to the modeling of the
mean sending time $\tau_i=\langle t_i^\mathrm{send} \rangle$:
\begin{eqnarray}
\label{eq:muout}
  \tau_i 
    &=&  1 + 
         \sum_{j_{1}\in\mathcal{N}_i^{in}} p_{j_1}(n_{j_{1}} \ge 1) +
         \sum_{j_{2}\in\mathcal{N}(\mathcal{N}_i^{in})
               \backslash\mathcal{N}_i^{in}}
         p_{j_2}(n_{j_{2}} \ge 1) 
         \sum_{j_{1}\in\mathcal{N}_i^{in}}
         \frac{B_{j_{2}\leftrightarrow j_{1}}}{2B_{j_{2}}}
         \nonumber \\
    &=&  1 + 
         \sum_{j_{1}\in\mathcal{N}_i^{in}} 
         \frac{\mu B_{j_1}}{(N-1)} \tau_{j_1} 
         + \sum_{j_2\in\mathcal{N}(\mathcal{N}_i^{in})
                 \backslash\mathcal{N}_i^{in}}
         \frac{\mu B_{j_2}}{(N-1)} \tau_{j_2} 
         \sum_{j_{1}\in\mathcal{N}_i^{in}}
         \frac{B_{j_2\leftrightarrow j_1}}{2B_{j_2}}
         \; .
\end{eqnarray}
The first sum in the first line represents those one-hop neighbors 
$j_1$, which also want to transmit a packet at the same time. Also 
two-hop neighbors $j_2$ contribute, once they have a packet to transmit 
to a one-hop neighbor, which would then \MAC-block node $i$; this is 
described by the last term of the first line. For the second step, Eqs.\
(\ref{eq:nigeone}) and (\ref{eq:muin}) have been used, leading to $N$ 
coupled linear inhomogeneous equations for the sending times, which are 
then solved numerically for a given network realization. Note, that the
expression (\ref{eq:muout}) overestimates the actual sending time to 
some extend, because one- and two-hop neighbors might have already been
blocked by previously assigned one-hop transmissions.

Upon putting Eqs.\ (\ref{eq:littlelaw}), (\ref{eq:avgni}), 
(\ref{eq:muin}) and (\ref{eq:muout}) together, an analytic estimate
of the average end-to-end time delay can finally be given. It is 
interesting to mention two limiting cases. In the limit $\mu\to 0$ of 
very small packet creation rates, the estimated
sending times (\ref{eq:muout}) converge to $\tau_i\to 1$, so that
the average end-to-end time delay becomes
\begin{equation}
\label{eq:limit1}
  \lim_{\mu\to 0} \langle t_{e2e} \rangle 
    =  \frac{1}{\mu N} \sum_i \mu_i^\mathrm{in}
    =  D
       \; ,
\end{equation}
where the sum rule $\langle B_i \rangle = (N-1) D$ has been used to 
express the mean node inbetweeness in terms of the network diameter
\cite{kra04}. This shows that for weak data traffic loads the network 
diameter determines the end-to-end time delay. In the other limit, 
$\mu\to\mu_{crit}$, the average end-to-end time delay is dominated by 
the most critical node. This node $i$ is the first, for which the in- 
and out-flux rates become identical and which determines the critical
packet creation rate $\mu_{crit} = (N-1)/(\sup_j B_j\tau_j)$. The sum
over all nodes $\sum_j n_j \approx n_i$ breaks down and we arrive at
\begin{eqnarray}
\label{eq:limit2}
  \lim_{\mu\to\mu_{crit}} \langle t_{e2e} \rangle
    =  \frac{1}{\mu N}
       \frac{1}{1-\frac{\mu B_i \tau_i(\mu)}{N-1}}
    \sim
       \frac{1}{\mu_{crit}-\mu}
       \; .
\end{eqnarray}
In the last step we have made use of the fact that $\tau_i(\mu)$ is
bounded from above by the magnitude of node $i$'s one- and two-hop 
neighborhood. As the packet creation rate approaches $\mu_{crit}$, the 
number of packets within the network explodes to infinity. The critical 
exponent turns out to be $1$.  

For the const-$P$ and minimum-node-degree networks, the analytic 
estimate of the average end-to-end time delay is shown in Fig.\ 
\ref{fig:te2e} and compared to the respective results obtained from the 
generic data traffic simulations. Note that both the analytic as well as
the simulation estimate have been averaged over a large sample of 
network realizations. For $\mu$ safely below $\mu_{crit}$ a good 
agreement is found. Since the analytic estimate and the generic 
data-traffic simulation produce a slightly different $\mu_{crit}$, 
divergence of $t_{e2e}$ sets in at different $\mu$, so that the quality 
of the comparison declines for $\mu$ close to $\mu_{crit}$.

\subsection{End-to-end time delay III: scalability}
\label{subsec:threeC}

\begin{figure}
\begin{centering}
\noindent
\epsfig{file=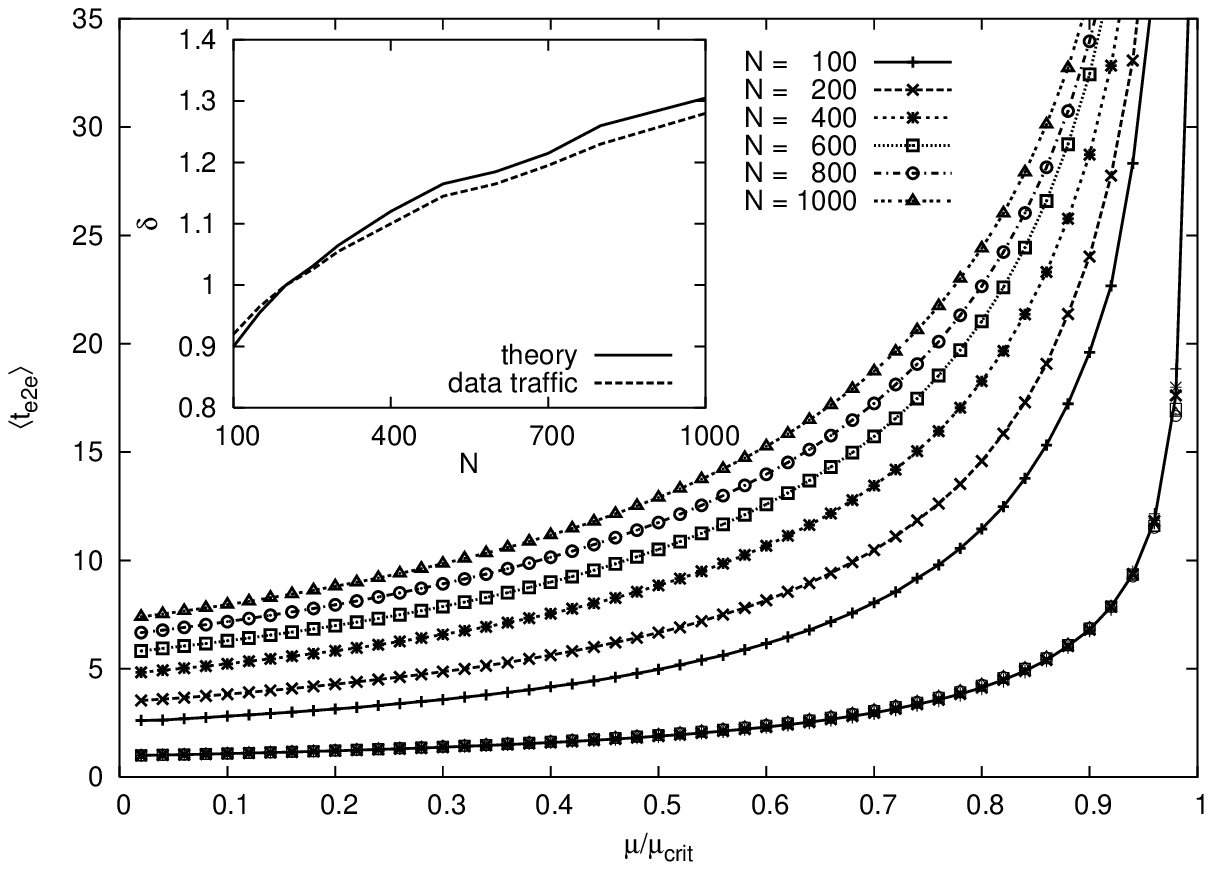,width=12cm}
\epsfig{file=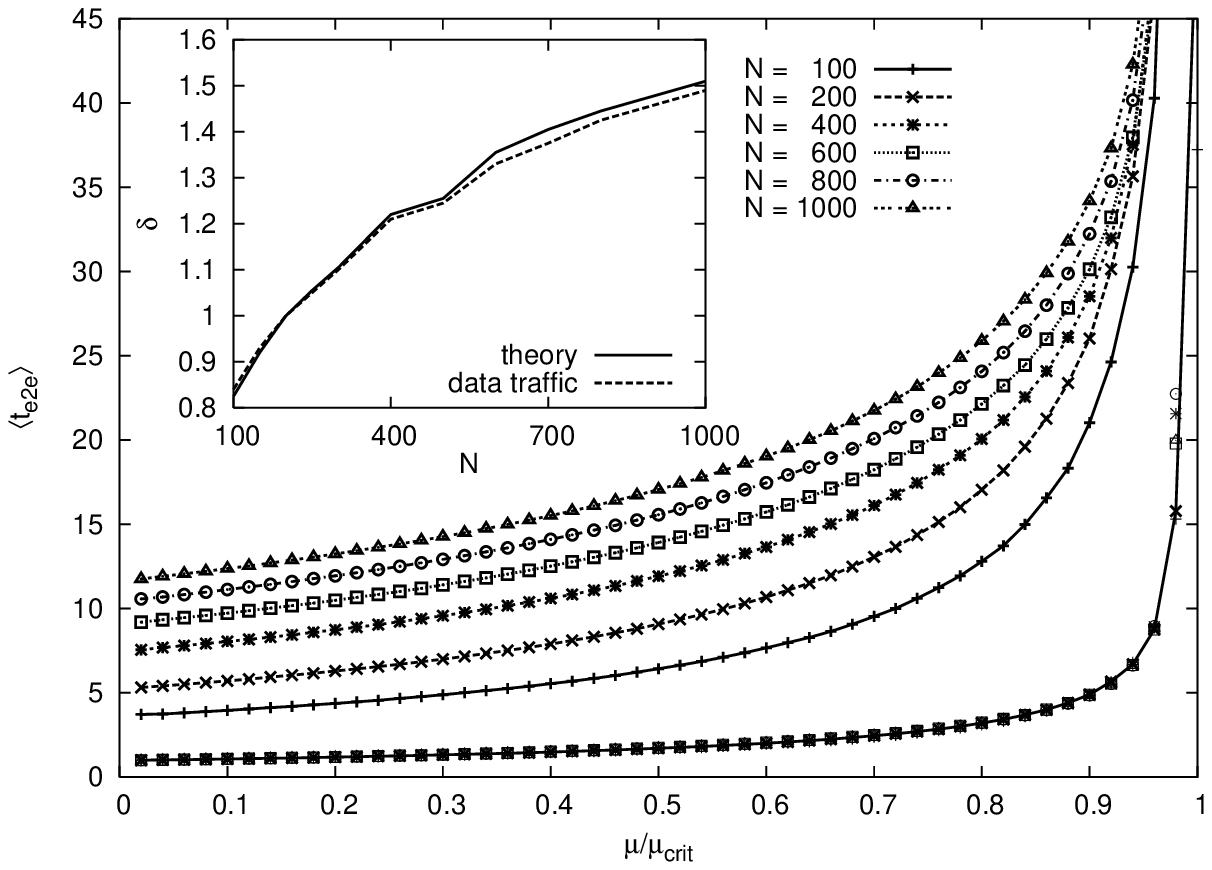,width=12cm}
\caption{
Average end-to-end time delay as a function of $\mu/\mu_{crit}$
for various sizes of (top) const-$P$ and (bottom) minimum-node-degree
networks. The lowest curve represents the curve collapse 
(\ref{eq:te2escaling}) with $N_0=200$; the exponent $\delta$ is shown 
in the inset figure.
}
\label{fig:te2escaling}
\end{centering}
\end{figure}

Instead of comparing the load-dependent end-to-end time delay between
the two network models of fixed size, a comparison within a given
network model, but of varying size allows to address the scalability
issue. Fig.\ \ref{fig:te2escaling} illustrates  
$\langle t_{e2e} \rangle$ as a function of $\mu/\mu_{crit}$ for various 
sizes $N\geq 100$ of const-$P$ and minimum-node-degree networks. For 
each $\mu/\mu_{crit}$ and fixed $N$ generic data traffic simulations 
have been run with a sample of $100$ network realizations. Note also, 
that for each realization the critical packet creation rate fluctuates 
to some small degree. For a fixed $\mu/\mu_{crit}$ the end-to-end time 
delay increases with network size. In the limit $\mu\ll\mu_{crit}$ this 
increase roughly scales as $\sqrt{N}$, which is in accordance with
$t_{e2e}(\mu=0|N)=D(N)\sim\sqrt{N}$, reflecting the scaling behavior
of the network diameter.

In order to make scalability statements also for 
$0<\mu/\mu_{crit}<1$, the following scaling ansatz is proposed:
\begin{equation}
\label{eq:te2escaling}
  \left(
  \frac{ t_{e2e}(\mu/\mu_{crit}|N) }{ t_{e2e}(0|N) }
  \right)^\delta
    =  \frac{ t_{e2e}(\mu/\mu_{crit}|N_0) }{ t_{e2e}(0|N_0) }
       \; .
\end{equation}
$N_0$ is some reference network size. An exponent $\delta=1$ would
imply that the relative increase of the end-to-end time delay with 
respect to relative network load is independent of the network size.
For $\delta>1$, the relative increase would decrease with network size 
and for $\delta<1$ it would be the other way around. In fact, the 
scaling ansatz (\ref{eq:te2escaling}) leads to an excellent curve 
collapse, also shown as the lowest curve in Fig.\
\ref{fig:te2escaling}(top)+(bottom): the solid curve corresponds 
to the right-hand side of (\ref{eq:te2escaling}) with $N_0=200$ and
the collapsing symboled curves correspond to the left-hand side of
(\ref{eq:te2escaling}). The fitted exponent $\delta$ is shown in the
inset figure and reveals an $N$-dependence of the form
$\delta(N)=(N/N_0)^\beta$. For const-$P$ networks with
$k_\mathrm{target} = 24$ and minimum-node-degree networks with
$k_\mathrm{min}=8$ we find $\beta=0.11$ and $0.25$, respectively. This 
outcome shows, that from the perspective of the relative end-to-end time 
delay minimum-node-degree networks scale better with increasing network 
size than const-$P$ networks.

Focusing on the size-dependence of the network models, the analytic 
estimate of the average end-to-end time delay also confirms the scaling 
ansatz (\ref{eq:te2escaling}) with $\delta(N)=(N/N_0)^\beta$; see insets 
of Fig.\ \ref{fig:te2escaling}(top)+(bottom). For minimum-node-degree 
networks the found value $\beta=0.25$ perfectly matches the outcome from 
the generic data traffic simulations, whereas for const-$P$ networks a 
small discrepancy remains between the theoretically found $\beta=0.15$ 
and its simulation counterpart $0.11$.

\subsection{Single-node Correlation time: simulation results}
\label{subsec:threeD}

Another reason for the emergence of congestion is that the queue length
at specific nodes may fluctuate strongly around its mean. For sure, the
occurrence of $n_i\gg\langle{n_i}\rangle$ with non-negligible probability
enhances congestion. More than this, it is also the enhanced time it 
takes in such cases to relax back to the mean. A long relaxation time 
would mean that the specific node as well as its surrounding part of the
network stay in a congested state for quite some time. A straightforward 
measure of such a relaxation time is given by the first moment 
$\langle n(t+\Delta t)|n(t) \rangle$ of the conditional probability 
$p(n(t+\Delta t)|n(t))$ to have a queue length $n(t+\Delta t)$ at time 
$t+\Delta t$ given $n(t)$ packets at time $t$. It is related to the 
correlation function $r(\Delta t)$ by averaging over all possible 
$n(t)$:
\begin{equation}
\label{eq:corrFCT}
  r(\Delta t)
    =  \frac{\langle n(t+\Delta t) \; n(t) \rangle}{\langle n \rangle^2}
    =  \frac{1}{\langle n \rangle^2}
       \sum_{n(t)} \langle n(t+\Delta t)|n(t) \rangle \; n(t) \; p(n(t)) 
       \; .
\end{equation}
Thus, the correlation function can also be seen as an averaged measure 
of relaxation times. 

\begin{figure}[t]
\begin{center}
\epsfig{file=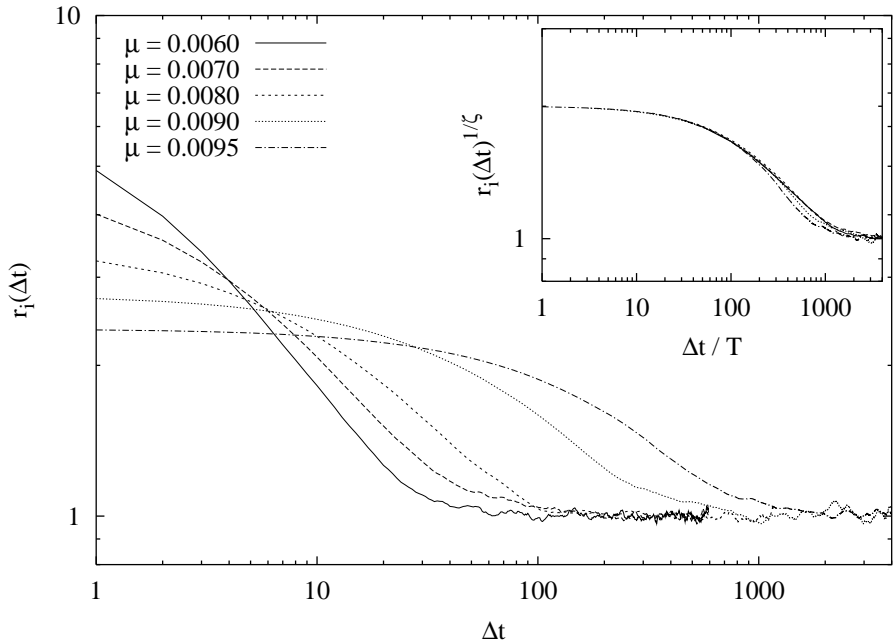,width=0.9\textwidth}
\caption{
Single-node temporal correlation $r_{i}(\Delta t)$ for the most-critical 
node of a const-$P$ reference network with $N=100$, obtained from a 
simulation covering $5\cdot10^5$ time steps. Different line types 
correspond to different packets creation rates, all below 
$\mu_\mathrm{crit}=0.0101$. The inset shows the curve collapse 
(\ref{eq:Tcollapse}) after appropriate rescaling.
}
\label{fig:2pointcorr_unscaled.constP24nodes100macB.node39}
\end{center}
\end{figure}

Fig.\ \ref{fig:2pointcorr_unscaled.constP24nodes100macB.node39}
illustrates the sampled single-node temporal correlation function
$r_i(\Delta t)$ for a characteristic node in a typical const-$P$
network. The convergence of $r_i(\Delta t)$ to $1$ for large time
differences indicates that correlations between queue lengths $n_i(t)$
and $n_i(t+\Delta t)$ no longer exist for large $\Delta t$. As the
packet creation rate grows, this decorrelation is shifted towards
larger time differences.

\begin{figure}[t]
\begin{center}
\epsfig{file=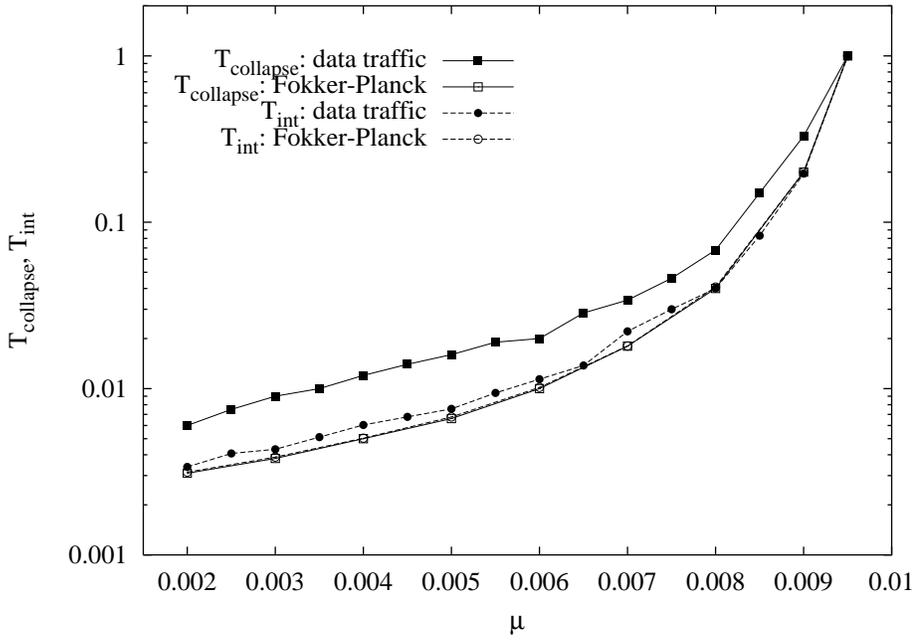,width=0.9\textwidth}
\caption{
Time scales $T_\mathrm{collapse}$ and $T_\mathrm{int}$ as a function of
the packet creation rate. The same const-$P$ reference network with 
$N=100$ has been used as in the previous Figure. The results directly 
obtained from a data traffic simulation are shown with filled symbols.
The curves with open symbols have been derived from Eq.\
(\ref{eq:FPfinalCorr2}) with in- and out-flux rates sampled from data
traffic simulations. Note that the two time scales have been normalized
such that $T_\mathrm{collapse/int}=1$ for $\mu=0.0095$.
}
\label{fig:2pointcorr_scaling.constP24nodes100macB.node39}
\end{center}
\end{figure}

A non-standard way to extract characteristic time scales results from 
the observation, that the various curves of Fig.\ 
\ref{fig:2pointcorr_unscaled.constP24nodes100macB.node39} for differing
packet creation rates all appear to have a similar functional form. This
motivates the curve collapse
\begin{equation}
\label{eq:Tcollapse}
  r_i(\Delta t) 
    = \left[ f \left( 
      \frac{\Delta t}{T_\mathrm{collapse}(\mu)} 
      \right) \right]^\zeta 
      \; ,
\end{equation}
where the condition 
$r_i(\Delta t{=}0) 
 = \langle n_i^2(t) \rangle / \langle n_i(t) \rangle^2$
fixes the exponent to
$\zeta 
 = \ln ( \langle n_i^2(t) \rangle / \langle n_i(t) \rangle^2 ) 
   / \ln f(0)$;
for later convenience we choose $f(0) = 2$. A suitable tuning of 
$T_\mathrm{collapse}(\mu)$ then leads to a perfect curve collapse; see 
the inset of Fig.\ 
\ref{fig:2pointcorr_unscaled.constP24nodes100macB.node39}. The extracted
time scale is shown in Fig.\ 
\ref{fig:2pointcorr_scaling.constP24nodes100macB.node39} as a function
of the packet creation rate. 
--
Another, now standard way to extract a characteristic time scale uses 
\begin{equation}
\label{eq:Tint}
  \int_0^\infty \left( r(t)^{1/\zeta} - 1 \right) dt
    =  \left( r(0)^{1/\zeta} - 1 \right) T_\mathrm{int}
\end{equation}
to define an integral time scale $T_\mathrm{int}$. Its dependence on the
packet creation rate is also shown in Fig.\ 
\ref{fig:2pointcorr_scaling.constP24nodes100macB.node39}.

\subsection{Single-node Correlation time: analytic estimate}
\label{subsec:threeE}
 
We will now give some semi-analytic understanding of the simulation 
findings of the previous Subsection. In the rearranged form,
\begin{eqnarray}
\label{eq:FPrate2}
  \lefteqn{ p(n,t+1) - p(n,t) }
  \nonumber \\
    &=&  \frac {\left(\mu^{\rm out}-\mu^{\rm in}\right)}{2} \,
         \left[ p(n+1,t) - p(n-1,t)\right]
         \nonumber \\
    & &  + \frac {\left(\mu^{\rm out}+\mu^{\rm in}\right)}{2} \,
         \left[ p(n+1,t) - 2p(n,t) + p(n-1,t)\right]
         \; , 
\end{eqnarray}
the rate equation (\ref{eq:bufferDistr1}) transforms into a 
Fokker-Planck equation
\begin{equation}
\label{eq:FPeq}
  \frac{\partial p(n,t)}{\partial t} 
    =  - \left(\mu^\mathrm{in}{-}\mu^\mathrm{out}\right)
       \frac{\partial p(n,t)}{\partial n} 
       + \frac{\left(\mu^\mathrm{in}{+}\mu^\mathrm{out}\right)}{2} 
       \frac{\partial^2 p(n,t)}{\partial n^2}
       \; .
\end{equation}
The invoked continuum limit from discrete to continuous $n$ is justified
for long queue lengths $\langle n \rangle \gg 1$, which is the case when 
the in- and out-flux rates $\mu^\mathrm{in} \approx \mu^\mathrm{out}$ 
are almost the same. The latter also determine the drift and diffusion 
coefficients 
\begin{equation}
\label{eq:DriftDiff}
  -\gamma
    =  \mu^\mathrm{in}{-}\mu^\mathrm{out}
       \; , \qquad\qquad
  D
    =  \mu^\mathrm{in}{+}\mu^\mathrm{out}
         \; ,
\end{equation}
which are both constant. Note, that a minus sign has been introduced in 
the definition of the drift coefficient, which guarantees $\gamma>0$ for 
the subcritical traffic regime $\mu^\mathrm{in}<\mu^\mathrm{out}$.

The solution of this Fokker-Planck equation with given initial condition 
$n(t{=}0)$ and reflecting boundaries at $n=0$ and $\infty$ requires an 
expansion into eigenfunctions \cite{gar83}. The eigenfunction method
also allows a direct calculation of the correlation function. This 
calculation is rather lengthy, but straightforward. Details are given in 
Ref.\ \cite{gla03b}. Here, we state only the final result,
\begin{eqnarray}
\label{eq:FPfinalCorr2}
  \langle n(t) \, n(0) \rangle 
    &=&  \frac{1}{4} \left( \frac{D}{\gamma} \right)^2 
         + \frac{4}{\pi} \left( \frac{D}{\gamma} \right)^2 
         e^{-\frac{\gamma^{2}}{2D}t} \, 
         \left[ \frac{\pi}{16} \, M\!\left( 
         \frac{3}{2},-\frac{1}{2},\frac{\gamma^2}{2D}t
         \right) \right. 
         \nonumber \\
    & &  \left. \quad \quad + \frac {2}{3} \sqrt{\pi} 
         \left( \frac{\gamma^2}{2D}t \right)^{3/2} 
         M\!\left( 3,\frac{5}{2},\frac{\gamma^2}{2D}t \right) \right] 
         \; ,
\end{eqnarray}
expressed in terms of confluent hypergeometric functions \cite{abr72}. 
It comes with the property 
$\langle n^2(0) \rangle = 2 \langle n(0) \rangle^2$, which is in 
agreement with the simulation results of the previous Subsection for 
packet creation rates close to the critical one (see again Fig.\ 
\ref{fig:2pointcorr_unscaled.constP24nodes100macB.node39}), where
$\langle n \rangle \gg 1$. It also explains the choice $f(0)=2$ 
introduced for the curve collapse (\ref{eq:Tcollapse}).

\begin{figure}[t]
\begin{center}
\epsfig{file=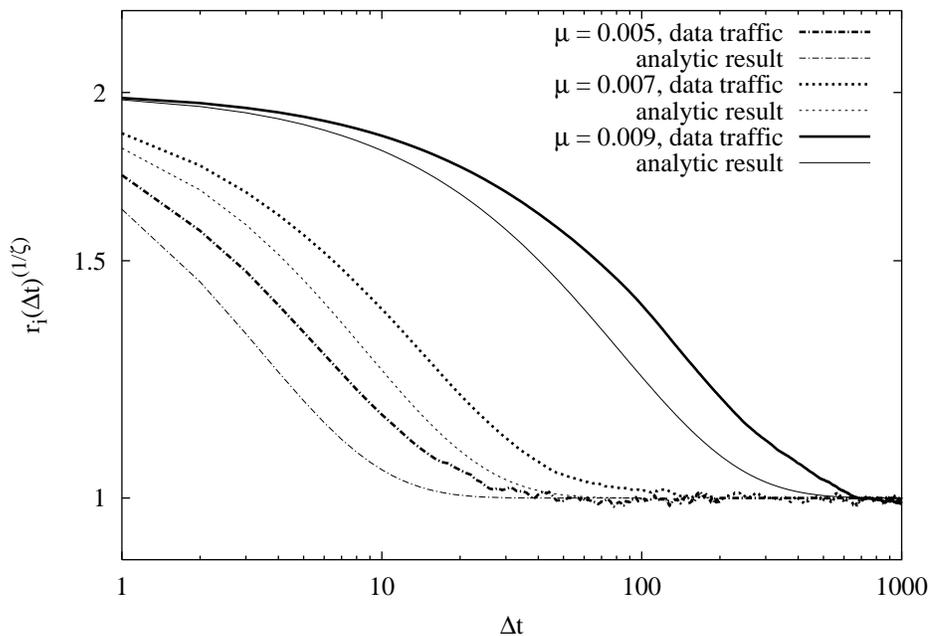,width=0.9\textwidth}
\caption{
Comparison of the rescaled correlation functions 
$(r_i(\Delta t))^{1/\xi}$ of Fig.\ 
\ref{fig:2pointcorr_unscaled.constP24nodes100macB.node39} with the 
analytic expression (\ref{eq:FPfinalCorr2}). For the latter, the in-
and out-flux rates have been sampled from the same generic data traffic 
simulations, the same network realization and the same node, which have
been used for the former.
}
\label{fig:comparisoncorrfct}
\end{center}
\end{figure}

The expression (\ref{eq:FPfinalCorr2}) only depends on the in- and 
out-flux rates (\ref{eq:DriftDiff}), otherwise its functional form is 
fixed. Upon taking the sampled $\mu_i^\mathrm{in}$ and 
$\mu_i^\mathrm{out}$ from the generic data traffic simulation,  
expression (\ref{eq:FPfinalCorr2}) can be compared with the directly 
sampled correlation functions; this is done in Fig.\ 
\ref{fig:comparisoncorrfct}. For all packet creation rates, the
decorrelation time obtained via (\ref{eq:FPfinalCorr2}) is 
systematically somewhat larger than for the directly sampled correlation 
functions. A possible reason for this discrepancy might be the invoked
rescaling $(r_i(\Delta t))^{1/\xi}$ of the latter. However, the 
functional form looks the same and also the respective time scales
$T_\mathrm{collapse}$ and $T_\mathrm{int}$ as a function of the packet
creation rates show a good agreement with the previously obtained 
results of Fig.\ 
\ref{fig:2pointcorr_scaling.constP24nodes100macB.node39}. The overall 
good correspondence between the simulation and the semi-analytic 
results more or less explains the increase of the correlation time 
scales with growing packet creation rate as an inherent feature of the 
underlying single-node queuing behavior.

\section{Routing \& congestion control}
\label{sec:four}

>From the previous Section on the generic data traffic with shortest-path
routing we have learned several things. It is the most critical node,
which gets overloaded first among all other nodes and which determines
the critical packet creation rate of the overall network. This limits 
the network's e2e-throughput capacity to $\mu_\mathrm{crit} N$. 
Moreover, for network loads close to $\mu_\mathrm{crit}$ a good fraction 
of nodes belonging to the greater neighborhood of the critical one will 
also come with large queue lengths $n_i \gg 1$. This congestion cluster 
then gives rise to large average e2e-time delays and to large 
fluctuations of the latter, which in turn result in large relaxation 
times. It is the goal of any routing \& congestion control to avoid such 
congested network areas and to detour the data traffic around. Such 
actions are likely to be rewarded with an increase of the e2e-throughput 
capacity, a decrease of the average e2e-time delay as well as its 
fluctuations and the related relaxation times. In this Section we 
discuss three different routing \& congestion controls of increasing 
sophistication. The first one exploits the degeneracy of shortest 
e2e-routes. The other two approaches modify the distance metric to 
include each node's congestion state and adapt the routing decisions 
according to updated cost estimates, which are locally exchanged with 
every \MAC-blocking.

\subsection{Simple congestion control with degenerated shortest paths}
\label{subsec:fourA}

The shortest-path (SP) routing used in Sect.\ \ref{sec:three} does not
take advantage of the route degeneracy between an arbitrary sender and a 
final receiver. Randomly choosing one out of several degenerate shortest 
routes for each new packet will already give some relief to the most 
congested nodes. However, a bias on the actual congestion state would do 
even better. A simple extension in this direction is shortest-path 
shortest-queue (SPSQ) routing. Instead of randomly choosing one next-hop 
neighbor out of the several degenerate shortest routes that specific 
node is picked which in addition has the shortest queue length in its 
buffer for this very moment. If more than one node qualifies, one of 
them is again picked randomly. Note, that in order to make such a 
routing decision, the forwarding node needs to have information from its 
neighbors about their congestion state. A very elegant way to provide
this information without sending additional control packets is to 
include it into the \MAC-blockings, in which the neighbors had been 
actively involved during previous one-hop transmissions.

\begin{figure}[t]
\begin{center}
\epsfig{file=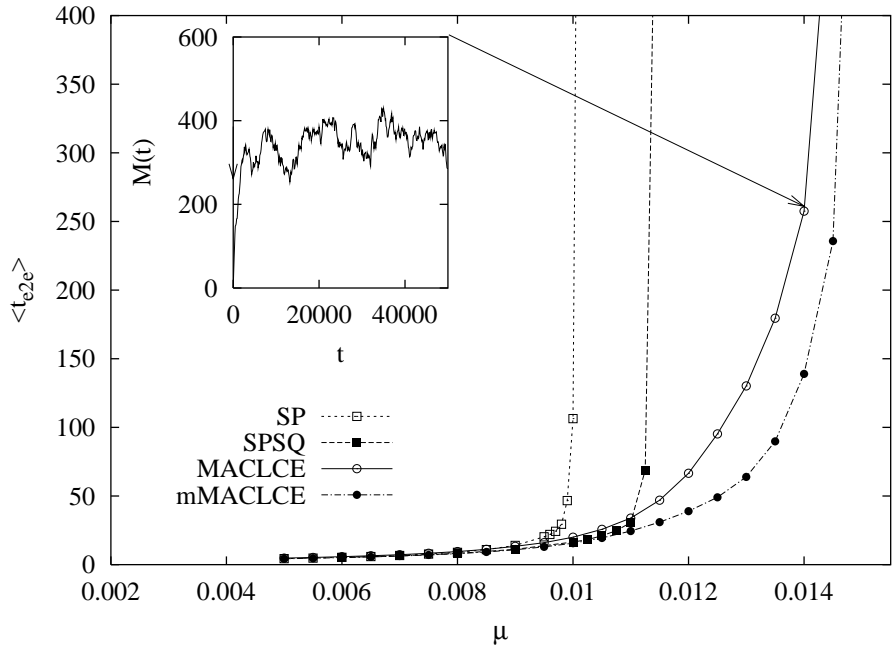,width=0.9\textwidth}
\caption{
Average end-to-end time delay as a function of the packet creation rate
for various routing \& congestion controls: 
(SP) shortest-path, 
(SPSQ) shortest-path-shortest-queue, 
(MACLCE) \MAC-distributed lowest-cost-estimate with $\nu=0$, and
(mMACLCE) memory-based \MAC-distributed lowest-cost-estimate with 
$\nu=0.65$.
Respective generic data traffic simulations have been run on an 
identical const-$P$ network realization of size $N=100$.
}
\label{fig:constP24nodes100macB.compare_meandelay}
\end{center}
\end{figure}

\begin{figure}
\begin{center}
\epsfig{file=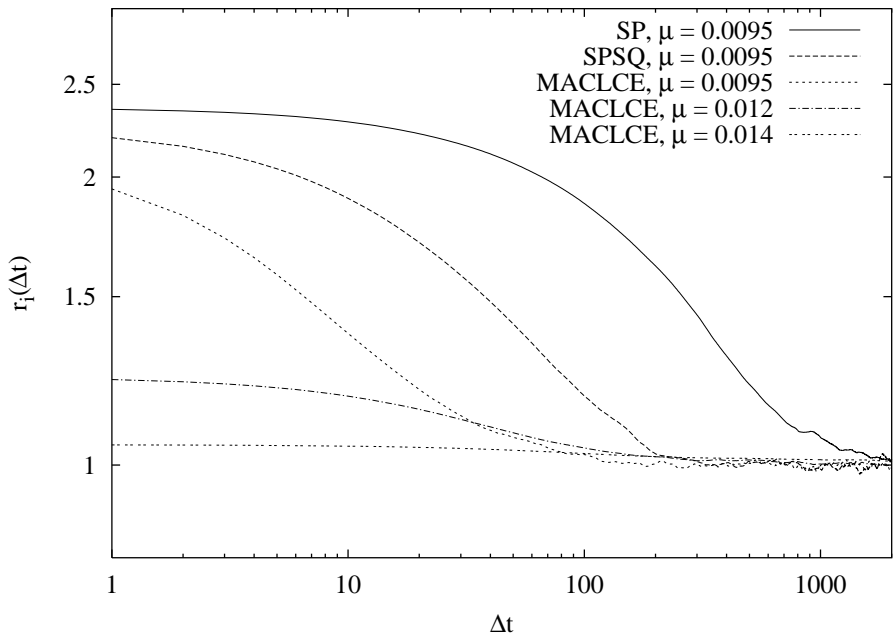,width=0.9\textwidth}
\caption{
Single-node correlation function $r_{i}(\Delta t)$ for various routing \& 
congestion controls and packet creation rates.
Respective generic data traffic simulations have been run on an identical 
const-$P$ network realization of size $N=100$ and also the picked node
has been the same for all cases.
}
\label{fig:constP24nodes100macBpcs1.adaptive_queue.2point}
\end{center}
\end{figure}

As can be seen from Figs.\ 
\ref{fig:constP24nodes100macB.compare_meandelay}, 
\ref{fig:constP24nodes100macBpcs1.adaptive_delayhisto} and 
\ref{fig:constP24nodes100macBpcs1.adaptive_queue.2point}, the simple 
SPSQ routing \& congestion control already leads to some noticeable 
improvements. The critical packet creation rate, where the average 
e2e-time delay diverges, increases. The tail of the e2e-time-delay 
distribution is suppressed to some extend. Last but not least, also the 
correlation time scales have become smaller.

\subsection{Congestion control with instantaneous adaptive routing}
\label{subsec:fourB}

If a node would have the complete information on the instantaneous 
congestion state of all other nodes belonging to the network, then it 
could determine the shortest route to the intended final recipient with
a modified metric, which does not only take the hop distance into 
account, but also counts the queue lengths of all inbetween nodes. 
However, first of all the node does not have this information, and even
if it would, then the shortest path at decision time needs not be equal
to the shortest path for delivery due to the always changing congestion 
state of the network during delivery time $t_{e2e}$. Instead, the node 
could try to get some sort of cost estimate it takes to send the packet 
to the final destination via this or that neighbor, and to constantly 
update these estimates. This idea is already known as asynchronous 
vector distance routing \cite{lit93} and has found its way into Internet
routing at the autonomous-system level. Next, we give an outline of this 
approach, modified and tuned to the specificity of wireless multihop ad 
hoc communication.

Pick a node $i$ that has a packet to forward to the final destination
$f$. Node $i$ has to decide to which neighbor $j\in\mathcal{N}_i$ it is
going to forward the packet. For each of them it has a cost estimate
$W_{if,j}$. It chooses node $j_\mathrm{min}$, providing the lowest cost
estimate 
$W_{if,j_\mathrm{min}} = \min_{j\in\mathcal{N}_i} W_{if,j} = W_{if}$.
Before $i$ starts transmitting its packet to $j_\mathrm{min}$, it has 
to \MAC-block its neighbors. Hand in hand with the blocking signal it
tells them about the minimum cost $W_{if}$ and its future queue length
$n_i-1$. While then being blocked, those neighbors have enough time to
process this information and update their estimated cost 
\begin{equation}
\label{eq:instantupdate}
  W_{jf,i} 
    \leftarrow  w_{ji} + W_{if}
\end{equation}
to send a packet via $i$ to $f$ during a future time step. $w_{ji}$ is 
the cost to send a packet from $j$ to its neighbor $i$. Since the queue 
length $n_i-1$ is a reasonable measure of how busy node $i$ is
\cite{lit93,heu98}, we set 
\begin{equation}
\label{eq:processingsender}
  w_{ji}  =  (n_i-1) + 1
             \; ;
\end{equation}
the one at the end takes care of the hop-distance between the two 
nodes. As the intended receiver of $i$'s to-be-transmitted packet, node
$j_\mathrm{min}$ also blocks its neighbors 
$k\in\mathcal{N}_{j_\mathrm{min}}$ and takes this chance to inform 
those about its estimated cost $W_{j_\mathrm{min}f}$ and its future 
queue length $n_{j_\mathrm{min}}+1$. Those neighbors then process this 
information analogous to (\ref{eq:instantupdate}), but with the 
modification 
\begin{equation}
\label{eq:processingreceiver}
  w_{kj_\mathrm{min}}  =  (n_{j_\mathrm{min}}+1) + 1
\end{equation}
for the future link cost.

Updating the estimated costs after receipt of an extended \MAC-signal 
appears to be a natural thing for wireless multihop ad hoc 
communication. It couples the \MAC-layer and the routing layer in a 
very elegant way and gives rise to a distributive congestion control. 
Note that beyond \MAC\ no additional congestion control signals have to 
be exchanged. This is in contrast to other communication networks like 
the Internet, which first probe the congestion state with additional 
dedicated signals before they adapt to it.

A few more technical words are in order. The congestion information
distributed by the \MAC-signals may either include only those cost 
estimates belonging to the final destination of the currently 
transmitted packet, or it may include much more than this, namely the 
cost estimates to every single node of the network. Both cases 
certainly represent two extremes, the former focusing on keeping the
distributed signal small and the latter allowing for a faster spread
of the congestion information over the entire network. For the 
remainder of this Paper, we will only concentrate on the latter case.

Implementing the proposed congestion control for the generic data 
traffic simulations, requires to specify the initialization. The 
initial cost estimates have been set to
\begin{equation}
\label{eq:PathInitial}
  W_{if,j}
    =  \left\{
       \begin{array}{ll}
         1      & \qquad (f\in\mathcal{N}_i) \\
         0      & \qquad (i=f) \\
         \infty & \qquad (\mathrm{else})
       \end{array}
       \right. 
       \; .
\end{equation}
The simulations themselves then have shown that for not-too-small 
packet creation rates the number of time steps it takes to distribute 
the initial cost estimates over the entire network and to reach a 
kind of steady state is of the order of the network size $N$. 
Furthermore, test simulations have revealed that a  
relaxation of the deterministic lowest-cost-neighbor choice for 
forwarding the data packet, such that also the other higher-cost 
neighbors become eligible in some probabilistic form, always leads 
to a degradation in performance.

The simulations of the generic data traffic for the
routing \& congestion control with the \MAC-distributed lowest-cost 
estimates (MACLCE) have turned out to be quite time-consuming. For 
this reason we present results only for one realization of a 
constant-$P$ network with size $N=100$. The simulation run took 
$5\cdot10^{5}$ time steps. As a function of the packet creation rate 
the average e2e-time delay is illustrated in Fig.\ 
\ref{fig:constP24nodes100macB.compare_meandelay}. For moderate
packet creation rates the various routing \& congestion schemes SP,
SPSQ and MACLCE perform about equally well, but when it comes to the
critical regime, the control with \MAC-distributed lowest-cost 
estimates yields by far the largest critical packet creation rate.
Although more packets are present in the network, such a routing \&
congestion scheme is able to handle packet creation rates, exceeding
the maximum rate obtained with the classical shortest-path routing
by a factor of about $1.4$. The inset of this Figure shows the number 
of active packets $N_\mathrm{active}(t)$ for MACLCE at $\mu = 0.014$, 
indicated by the arrow. There is no linear dependence on time, which
is a clear sign that the network is still operating in the 
subcritical regime.

\begin{figure}[t]
\begin{center}
\epsfig{file=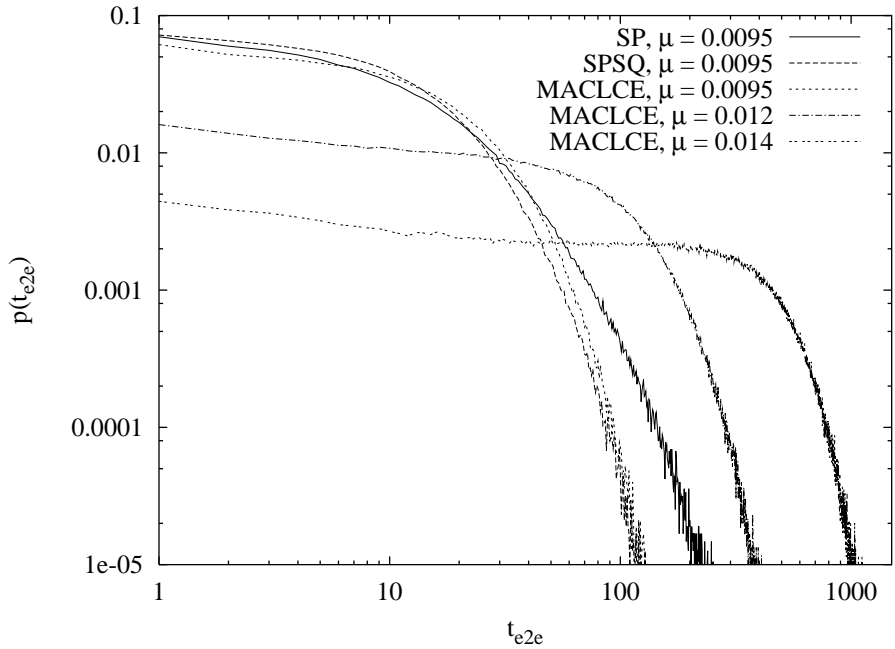,width=0.9\textwidth}
\caption{
Distribution of the end-to-end time delay for various routing \& 
congestion schemes and packet creation rates. For $\mu=0.0095$ the 
schemes shortest-path (SP), shortest-path-shortest-queue (SPSQ) and
\MAC-distributed-lowest-cost-estimates (MACLCE) are compared; for 
$\mu=0.012$ and $0.014$ only the MACLCE scheme is shown. Generic data 
traffic simulations have been run on an identical const-$P$ network 
realization of size $N=100$. 
}
\label{fig:constP24nodes100macBpcs1.adaptive_delayhisto} 
\end{center}
\end{figure}

Another goal for the new routing \& congestion algorithm has been to 
reduce the pronounced tail in the distribution $p(t_{e2e})$ of the 
end-to-end time delay. Fig.\ 
\ref{fig:constP24nodes100macBpcs1.adaptive_delayhisto} demonstrates
that for $\mu = 0.0095$, which is very close to the SP critical
packet creation rate, the MACLCE and SPSQ distributions are about
the same and, when compared to the SP distribution, come with a 
suppressed tail. Note, that log-scales have been used in Fig.\ 
\ref{fig:constP24nodes100macBpcs1.adaptive_delayhisto}.
For larger packet creation rates, where the SP and
SPSQ schemes are already in the supercritical regime, the routing \& 
congestion control with \MAC-distributed lowest-cost estimates
results in rather flat distributions, which come with a relative
sharp cutoff at larger end-to-end time delays. Of course, this cutoff
increases with the network load.

\begin{figure}[t]
\begin{center}
\epsfig{file=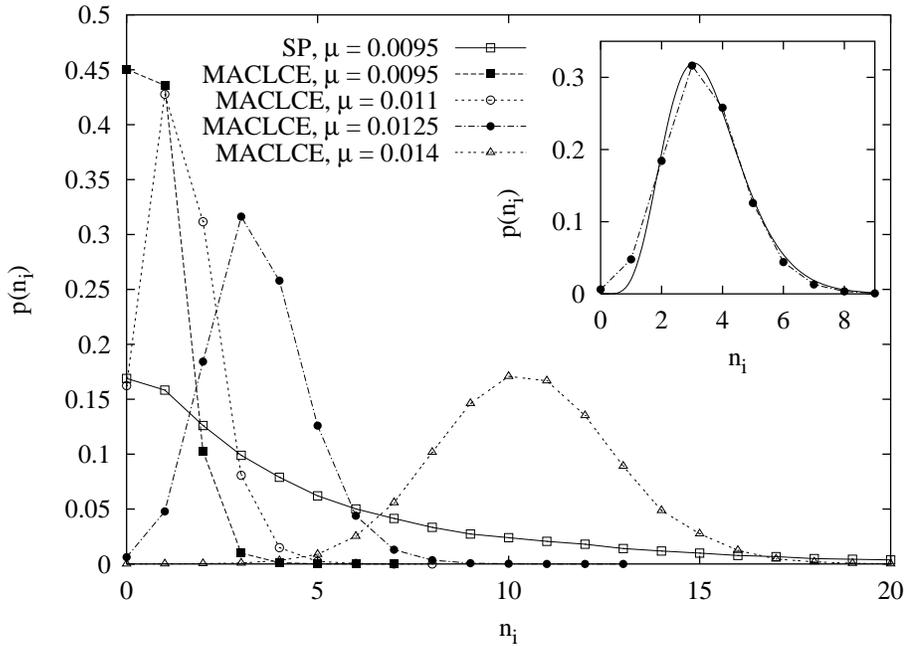,width=0.9\textwidth}
\caption{
Single-node distribution $p(n_{i})$ of the queue length for various 
routing \& congestion controls and packet creation rates.
Respective generic data traffic simulations have been run on an identical 
const-$P$ network realization of size $N=100$ and also the picked node
has been the same for all cases. For the MACLCE-curve with $\mu=0.0125$ 
the inset illustrates a fit with the Gamma-distribution 
(\ref{eq:gammaPDF}).
}
\label{fig:constP24nodes100macBpcs1.adaptive_queue.buffer}
\end{center}
\end{figure}

Compared to the SP routing significant changes can be observed in the 
distribution $p(n_i)$ for the single-node queue length; see Fig.\
\ref{fig:constP24nodes100macBpcs1.adaptive_queue.buffer}. At 
$\mu=0.0095$ the SP distribution is a broad exponential (see also 
bottom of Fig.\ \ref{fig:histo}), whereas the MACLCE distribution is
confined to basically $n_i \leq 3$. For larger packet creation rates
the MACLCE distributions become bell-shaped. Their mean and variance
increase with $\mu$. Good fits can be obtained with a two-parametric
continuous Gamma distribution
\begin{equation}
\label{eq:gammaPDF}
  p_\mathrm{gamma}(n) 
    =  \frac{1}{\Gamma(a) b^a} n^{a-1} e^{-\frac{n}{b}} 
       \; ;
\end{equation}
consult the inset Figure. For MACLCE-controlled networks close to  
$\mu^{\rm crit}$ this bell-shaped single-node queue-length 
distribution is a result of the flexible routing scheme, which adapts 
well to the current congestion state, and is key to their improved 
operational functionality.

A direct consequence of the bell-shaped distributions $p(n_i)$ is 
that $\langle n_i^2(t) \rangle \approx \langle n_i(t) \rangle^2$. 
This then keeps the correlation function (\ref{eq:corrFCT}) close to 
one, even for small $\Delta t$. Thus with the packet creation rate 
approaching its critical limit, the single-node temporal correlations 
are expected to be drastically reduced. The simulation results
illustrated in Fig.\ 
\ref{fig:constP24nodes100macBpcs1.adaptive_queue.2point}
confirm this view.

\subsection{Congestion control with memory-based adaptive routing}
\label{subsec:fourC}

So far the cost estimate $W_{if,j}$ of node $i$ is updated according to 
\eqref{eq:instantupdate} as soon as the neighboring node 
$j\in\mathcal{N}_i$ \MAC-reports its change of either $w_{ij}$ or 
$W_{jf}$. This immediate update might be too fast to be optimal. It could 
be wise to keep at least in parts some of the old cost estimate. 
Introducing a kind of memory parameter $0\leq\nu\leq 1$, a proposition for 
a modified update rule would be 
\begin{equation}
\label{eq:PathUpdateLearning}
  W_{if,j} 
    \leftarrow  \nu \, W_{if,j} + (1-\nu) \, [ \, w_{ij} + W_{jf} \,]
                \; .
\end{equation}
For $\nu>0$ a fraction of the old cost estimate is kept as a part of the 
new estimate. This modification includes a memory of formerly used routes 
that are only updated if significant changes in the network's congestion 
state have occurred. This approach, which is inspired by reinforcement 
learning \cite{lit93}, is known as Q-routing \cite{cho96}.

The simple extension (\ref{eq:PathUpdateLearning}) proved to cause a small 
but clearly measurable further improvement in network performance. As
indicated in Fig.\ \ref{fig:constP24nodes100macB.compare_meandelay} the 
memory-based mMACLCE scheme with $\nu=0.65$ increases the critical
packet-creation rate a little further when compared to the instantaneous
MACLCE scheme with $\nu=0$. It turns out that $\nu\approx 0.65$ is the 
optimal choice. As a function of $\nu$ the critical packet creation rate 
first increases for values rising from $\nu=0$ to $0.65$, takes its 
maximum at $\nu\approx 0.65$ and then decreases for values above. It is 
intuitive that in case of a strong memory ($\nu\approx 1$) the update rule 
collapses and only keeps very old and out-fashioned cost estimates, which 
are not suited to adapt to the always changing congestion state of the 
network.

\section{Conclusion and Outlook}
\label{sec:conclusion}

This Paper has focused on routing \& congestion control in wireless 
multihop ad hoc communication networks. Simulations with random data 
traffic have been accompanied with analytic estimates, whenever possible. 
The focus has first been on shortest-path routing, to understand certain 
data traffic characteristics like end-to-end time delay and correlation 
time scales, and to find fingerprints of congestion once the network is 
operating close to its critical load. A scaling law has been found for 
the average end-to-end time delay with respect to network size, which 
also revealed a dependence on the underlying network topology. In a 
second step and going beyond shortest-path routing, a distributive 
routing \& congestion control has been proposed, which couples the \MAC- 
and routing layer of wireless multihop ad hoc communication. Before 
one-hop forwarding a packet, the sending as well as receiving node 
\MAC-block their respective neighbors and distribute information about 
their congestion state and routing cost estimates, which the latter then 
use for updates. This distributive scheme turned out to be very 
efficient. Compared to shortest-path routing, the critical network load 
increased noticeably. Routes are constantly adapting to the prevailing 
congestion state of the network. With other words, routes selforganize 
themselves.

The proposed prototype routing \& congestion control needs of course
further testing and extensions. Other than simple random data traffic
has to be looked at, such as for example selfsimilar \cite{par00}, 
self-organized-critical \cite{val02} and spatially localized.
Congestion updates with different forms of cost metrics are important to 
investigate as well as the sparsity issue, i.e.\ which information is 
important to be distributed to other nodes and which is negligible. At 
the end, the biggest challenge is yet to come, to turn provably good 
ideas into real-life-functioning implementations. This is where physics 
and engineering should meet again.

\ack
W.\ K.\ acknowledges support from the Ernst von Siemens-Scholarship.

\newpage

\newpage

\end{document}